\documentclass[aps,pra,preprint,amsmath,amssymb]{revtex4}
\usepackage{graphicx,epsfig,epsf}
\usepackage{dcolumn}
\usepackage{bm}

\begin{document}
\newcommand{\ri}{ i}
\newcommand{\re}{ e}
\newcommand{\bx}{{\bf x}}
\newcommand{\bd}{{\bf d}}
\newcommand{\be}{{\bf e}}
\newcommand{\br}{{\bf r}}
\newcommand{\bk}{{\bf k}}
\newcommand{\bE}{{\bf E}}
\newcommand{\bI}{{\bf I}}
\newcommand{\bR}{{\bf R}}
\newcommand{\bZero}{{\bf 0}}
\newcommand{\bM}{{\bf M}}
\newcommand{\bn}{{\bf n}}
\newcommand{\bs}{{\bf s}}
\newcommand{\tbs}{\tilde{\bf s}}
\newcommand{\rSi}{{\rm Si}}
\newcommand{\beps}{\mbox{\boldmath{$\epsilon$}}}
\newcommand{\rg}{{\rm g}}
\newcommand{\tr}{{\rm tr}}
\newcommand{\xmax}{x_{\rm max}}
\newcommand{\ra}{{\rm a}}
\newcommand{\rx}{{\rm x}}
\newcommand{\rs}{{\rm s}}
\newcommand{\rP}{{\rm P}}
\newcommand{\up}{\uparrow}
\newcommand{\down}{\downarrow}
\newcommand{\hc}{H_{\rm cond}}
\newcommand{\kb}{k_{\rm B}}
\newcommand{\cI}{{\cal I}}
\newcommand{\tit}{\tilde{t}}
\newcommand{\cE}{{\cal E}}
\newcommand{\cC}{{\cal C}}
\newcommand{\Ubs}{U_{\rm BS}}
\newcommand{\qq}{{\bf ???}}
\newcommand*{\etal}{\textit{et al.}}
\def\vec#1{\mathbf{#1}}
\def\ket#1{|#1\rangle}
\def\bra#1{\langle#1|}
\def\keps{\mathbf{k}\boldsymbol{\varepsilon}}
\def\dm{\boldsymbol{\wp}}

\sloppy

\title{Spontaneous emission from a two--level atom tunneling in a
  double--well potential}
\author{Daniel Braun and John Martin }
\affiliation{Laboratoire de Physique Th\'eorique -- IRSAMC, CNRS,
  Universit\'e Paul Sabatier, Toulouse, FRANCE}

\centerline{\today}
\begin{abstract}
We study a two-level atom in a double--well potential coupled to a
continuum of electromagnetic modes (black body radiation in three
dimensions at zero absolute temperature). Internal and external
degrees of the atom couple due to recoil during emission of a
photon. We provide a full analysis of the problem in the long
wavelengths limit up to the border of the Lamb-Dicke regime,
including a study of the internal dynamics of the atom (spontaneous
emission), the tunneling motion, and the electric field of the
emitted photon. The tunneling process itself may or may not decohere
depending on the wavelength corresponding to the internal transition
compared to the distance between the two wells of the external
potential, as well as on the spontaneous emission rate compared to
the tunneling frequency. Interference fringes appear in the emitted
light from a tunneling atom, or an atom in a stationary coherent
superposition of its center--of--mass motion, if the wavelength is
comparable to the well separation, but only if the external state of the
atom is post-selected. \end{abstract}
\maketitle

\section{Introduction}
Young's double slit experiment, in which interference is observed from light
passing through two small slits or holes in a plate placed at a distance
comparable to the wavelength of the light, constitutes one of the
experiments at the base of quantum mechanics. Theory and experiment have been
refined over the years to the point that the two holes have been replaced by
two trapped atoms or ions which scatter incoming laser light
\cite{Lenz93,Eichmann93,Itano98,Agarwal02,Feagin06,Wickles06}.
With the
advance of the coherent control of the external degrees of freedom of
atoms (see
e.g.~\cite{Sebby06,Shin04,Haensel01,Treutlein06}), the realization of atom
interferometers \cite{Carnal91,Keith91,Hackermueller04} (see \cite{Miffre06}
for a recent review), and in
particular the
realization of Schr\"odinger cat states of the center--of--mass coordinate of
a single atom or ion \cite{Monroe96}, it is natural to ask if interference
could be observed in the light emitted from a {\em single} atom superposed
coherently in two different positions. A similar question was answered to the
negative in a paper by Cohen-Tannoudji {\em et al.} \cite{Cohen92} for the
case of scattering of light from a massive object brought into orthogonal
position states. The physical reason is clear: interference can only arise
if the
probe particle can distinguish the two locations of the target. But then the
probe particle must get entangled with the target. If the target is massive
its two orthogonal position states remain unaltered during scattering, and
therefore lead to
vanishing overlap of the scattered probe states after tracing out the
target. Later the scattering problem was
reconsidered for lighter targets, where it was shown that interference can
arise. In
particular, Rohrlich {\em et al.} \cite{Rohrlich06} analyzed the general
situation of the
scattering of two free particles, a probe with mass $m$ and a target with mass
$M$. Interference was predicted for the case of $m\simeq M$, and even
perfect visibility of interference fringes for $m=M$ in one dimension. \\

Similarly, Schomerus and coworkers \cite{Schomerus02} analyzed the
scattering of particles from a ``quantum obstacle'', an obstacle brought
into a coherent superposition of positions. An important difference to
\cite{Rohrlich06} lies in the fact that the target was supposed to be bound
in a double--well potential and to tunnel
coherently between the two wells with tunneling frequency
$\Delta$. For the case of one dimensional scattering, they showed that the
quantum obstacle leads to almost the same transmission resonances as two
fixed obstacles, if the kinetic energy $\epsilon$ of the incident particle
satisfies $\epsilon\ll \Delta$. In the opposite limit, interference can
still be recovered by post-selecting the elastic scattering channel.

Spontaneous emission is not the same as scattering, and it is {\em a priori}
unclear if these results apply to spontaneous emission alone as
well. Furthermore, the properties of the emitted light are only a small
part of the interesting physics that can arise, if internal and external
degrees of the tunneling atom are coupled. Indeed, one might ask, if
spontaneous emission itself (e.g.~the decay rates of the excited level)
changes, if the atom is brought into a coherent superposition of different
external states. Also, what happens with the tunneling motion? To what
extent does the emitted photon cause decoherence of the
external  degree of freedom? Spontaneous emission from a tunneling
two--level atom was
considered in  \cite{Japha96} for the case of the transmission of an
atom through a rectangular energy barrier in one dimension with the atom
coupled to a one
dimensional mode continuum. It was shown
that the recoil from photon emission can lift the atom over the tunnel
barrier and thus increase significantly the transmission.

In the present paper we examine spontaneous emission of a two--level
atom trapped in a double--well potential, where the atom interacts
with the full three dimensional continuum of electromagnetic modes
(the interaction with a single cavity mode was treated in
\cite{Martin07}). We carefully investigate the effective dynamics of
all three subsystems involved: the internal degree of freedom of the
atom, the tunneling motion, and the electric field created by the
emitted photon in a regime where the photon wavelength is at most
comparable to the well separation. The tunneling motion may suffer
decoherence from the emission of a photon from the atom, but this
depends, amongst other things, on the timing of the emission of the
photon. Interference in the light emitted from the atom is very
weak, but interference with perfect visibility of the fringes arises
if the external state of the atom is post-selected in the energy
basis.

\section{Model}
\subsection{Derivation of the Hamiltonian}
\begin{figure}
\begin{center}
\includegraphics[width=13cm]{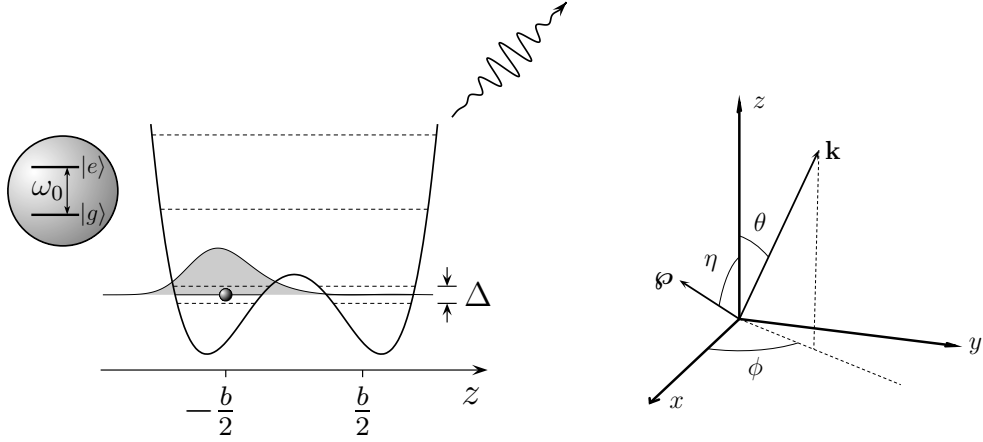}
\end{center}
\caption{Two-level atom in a double--well potential interacting with
a continuum of electromagnetic waves. Right panel: coordinate system used,
with the atom tunneling in $z$-direction, and the atomic dipole in the
$x-z$-plane} \label{doublewell}
\end{figure}

We consider a trapped two-level atom (with levels $|g\rangle$,
$|e\rangle$ of energy $\mp \hbar \omega_0/2$ respectively)
interacting with  traveling modes of the electromagnetic field as
illustrated in Fig.~\ref{doublewell}. The atom is assumed
to be tightly bound in the $x-y$ plane at the equilibrium position $x=y=0$
and to experience a symmetric double--well potential $V(z)$ along the
$z$ direction.
The Hamiltonian of this system is given by
\begin{equation}\label{tHamiltonian}
    H= H_{A}+H_{F}+H_{AF},
\end{equation}
where $H_{A}=H_{A}^{\mathrm{ex}}+H_{A}^{\mathrm{in}}$ denotes the
Hamiltonian of the trapped atom, $H_{F}$ is the Hamiltonian of the
free field and $H_{AF}$ is the interaction Hamiltonian describing
the atom--field interaction. Explicitly,
\begin{equation}\label{tHAex}
    H_{A}^{\mathrm{ex}}= \frac{p_z^2}{2M}+V(z),
\end{equation}
\begin{equation}\label{tHAin}
    H_{A}^{\mathrm{in}}=
    \frac{\hbar\omega_0}{2}\sigma_z^{\mathrm{in}},
\end{equation}
\begin{equation}\label{tHF}
    H_{F}= \hbar\sum_k\omega_k a_k^{\dagger}a_k,
\end{equation}
 and, in dipole approximation,
\begin{equation}\label{tHAF}
    H_{AF}= -\sum_k\mathbf{d.E}_k
\end{equation}
where $\mathbf{d}$ is the atomic dipole operator and
\begin{equation}\label{E}
    \mathbf{E}_k=
    {\cal E}_k\beps_k\left(a_k\re^{\ri \bk.
    \bR}+a_k^\dagger \re^{-\ri \bk.\bR}\right)
\end{equation}
the electric field operator, ${\cal E}_k=\sqrt{\frac{\hbar
\omega_k}{2\epsilon_0V}}$, $\epsilon_0$ is the permittivity of free
space, $V$ the electromagnetic mode quantization volume, $\beps_k$
the electric field polarization vector (normalized to length one),
$k$ stands for wave number $\bk=(k_x,k_y,k_z)$ and polarization
$\lambda=1,2$ of the electromagnetic modes with frequency $\omega_k=c|\bk|$ (where
$c$ is the speed of light in vacuum), and $\bR=(x,y,z)$ for the
center--of--mass position of the atom. Note that in
Eqs.~\eqref{tHAex}--\eqref{E} $z$ is still an operator, with $p_z$
its conjugate momentum for the atomic center-of-mass motion along
the $z$ axis; $M$ denotes the atomic mass, $\sigma_z^{\mathrm{in}} =
|e \rangle \langle e|-|g \rangle \langle g|$, and $a_k$
($a^{\dagger}_k$) the annihilation (creation) operator of mode $k$
of the radiation field.

In the following we will resort to the two-level approximation of
the motion in the external potential which amounts to taking into
account only the two lowest energy levels of the Hamiltonian
$H_A^{\mathrm{ex}}$. We denote by $\Delta$ the tunnel splitting,
i.e.\ the energy spacing between the two lowest energy states (the
symmetric $|-\rangle$ and antisymmetric $|+\rangle$ states) of the
double--well potential. Within this approximation, Hamiltonian (\ref{tHAex})
becomes
\begin{equation}\label{HAex}
    H_{A}^{\mathrm{ex}}= \frac{\hbar\Delta}{2}\sigma_z^{\mathrm{ex}}
\end{equation}
with $\sigma_z^{\mathrm{ex}} = |+ \rangle \langle +|-|- \rangle
\langle -|$.
We
can form states that are mainly concentrated in the left/right wells
by superposing the symmetric $|-\rangle$ and antisymmetric
$|+\rangle$ states,
\begin{equation}\label{LR}
\begin{aligned}
    |L\rangle &{}= \frac{|+\rangle-|-\rangle}{\sqrt{2}},\\
    |R\rangle &{}= \frac{|+\rangle+|-\rangle}{\sqrt{2}}.
\end{aligned}
\end{equation}
The position operator $z$ reads $z=b\,\sigma_x^{\mathrm{ex}}/2$ in the
two-level approximation, where
$\sigma_x^{\mathrm{ex}}=|-\rangle\langle+|+|+\rangle\langle-|=|R\rangle\langle
R|-|L\rangle\langle L|$, and $b/2=\langle +|z|-\rangle=\langle
R|z|R\rangle$ is the average $z$--position of the atom localized in
the right well [see Fig.~\ref{doublewell}]. 

The two-level approximation is justified if 
the higher vibrational energy levels are not populated during the
spontaneous emission process. This is well satisfied, if the recoil
energy $(\hbar\omega_0/c)^2/2M$ is much smaller than the difference
in energy to the next highest vibrational level in the external
potential (Lamb-Dicke regime). For an harmonic
potential, this implies
that the wavelength $\lambda$ of the emitted photon is much larger
than the extension of the ground state wave function. Here, we want
to reach the regime where $\lambda\sim b$ to observe interference in the
emitted light.
The localization of the states $|R\rangle$ and $|L\rangle$ compared
to $\lambda$ is measured by the parameter $\beta=\omega_0 b /c=2\pi
b/\lambda$. It turns out that at the border of the Lamb-Dicke regime
in a typical double--well potential, $\beta\sim 3$ (i.e.\
$\lambda\sim 2b$) still leads to a reasonable restriction of the
dynamics to the two lowest states [see Section~\ref{sec.exp} for
further details].

By expressing the dipole operator as $\mathbf{d}=\langle
e|\mathbf{d}|g\rangle\,\sigma_x^{\mathrm{in}}\equiv\wp\,\beps_d\sigma_x^{\mathrm{in}}$,
the interaction Hamiltonian~(\ref{tHAF}) can be cast in the form
\begin{equation}\label{HAF2}
    H_{AF}= \hbar\sum_k g_k\sigma_x^{\mathrm{in}}\left( a_k\re^{\ri \bk.
    \bR}+a_k^\dagger \re^{-\ri \bk.\bR}\right)
\end{equation}
with atom-field coupling strength $g_k=-\wp\,\beps_d.\beps_k\,{\cal
E}_k/\hbar$, dipole matrix element $\wp$, and the unit vector
$\beps_d$ in the direction of the vector of dipole matrix elements,
which we take without restriction of generality in the $x-z$-plane
with components $\beps_d=(\sin\eta,0,\cos\eta)$ (see
Fig.\ref{doublewell}). Furthermore, 
$\sigma_x^{\mathrm{in}}=\sigma_+^{\mathrm{in}}+\sigma_-^{\mathrm{in}}$,
$\sigma_-^{\mathrm{in}}=|g\rangle\langle e|$,
$\sigma_+^{\mathrm{in}}=|e\rangle\langle g|$,  and $\exp(\ri
\bk.\bR)=\cos\kappa+\ri \sin\kappa\,\sigma_x^{\mathrm{ex}}$ with
$\kappa=k_zb/2$. We will consider in the following the situation
where $\omega_0\gg \Delta$, and introduce the small parameter
$\delta\equiv\Delta/\omega_0\ll 1$. Experimentally, this is 
the most accessible situation (see Sec.~\ref{sec.exp}). A rotating
wave approximation is then in order, which leads to the interaction
Hamiltonian,
\begin{eqnarray} \label{HAFrw}
H_{AF}&=&\hbar\sum_k
g_k\Bigg(\cos\kappa\left(a_k\sigma_+^{\mathrm{in}}+a_k^\dagger\sigma_-^{\mathrm{in}}\right)
+\ri\sin\kappa\,\sigma_x^{\mathrm{ex}}
\left(a_k\sigma_+^{\mathrm{in}}-a_k^\dagger\sigma_-^{\mathrm{in}}\right) \Bigg)\,.
\end{eqnarray}
Note that in the case of $\Delta\sim\omega_0$ two additional terms
would have to be kept,
$i\sin\kappa(a_k\sigma_+^{\mathrm{ex}}\sigma_-^{\mathrm{in}}-
a_k^\dagger\sigma_-^{\mathrm{ex}}\sigma_+^{\mathrm{in}})$. Equation
(\ref{HAFrw}) makes clear that different electromagnetic waves will
act in quite different ways: waves with $\sin\kappa=0$ will only
interact with the internal degree of freedom, but leave the atom
position untouched. Indeed, these waves do not distinguish between
the left and the right well. Waves with $\sin\kappa\ne 0$, however,
will couple to both internal and external degrees of freedom at the
same time and can thus modify the tunneling behavior of the atom.

It is evident from Eq.~(\ref{HAFrw}) that the
tunneling motion can in principle reduce spontaneous emission: The
coupling constant of the second term in (\ref{HAFrw}) changes its
sign with the position of the atom in the double--well (``+'' in the
right ($z>0$) well, ``$-$'' in the left well), as $|R\rangle$ and
$|L\rangle$ are eigenstates of $\sigma_x^{\mathrm{ex}}$ with $\pm 1$
as eigenvalues. In the case of rapid tunneling motion, the sign of
that part of the Hamiltonian is therefore reverted, so is the
corresponding time evolution, and spontaneous emission is thus
reduced. However, reverting the time evolution has to happen on the time
scale of the dominating bath modes (i.e. $1/\omega_0$) in order to
give a significant effect. Based on Eq.~\eqref{HAFrw}, one may
expect at most a reduction by a factor 2 in the rate of spontaneous
emission for $\Delta\sim\omega_0$ as the first term in (\ref{HAFrw}) is
independent of the 
external degree of freedom of the atom. In the limit $\delta\ll 1$ which we
consider in this paper, the change of $\Gamma$ will turn out to be very
small. 

We will describe all dynamics in the interaction picture with the free
Hamiltonian $H_0=H_A^{\mathrm{in}}+H_A^{\mathrm{ex}}+H_F$. The corresponding time dependent
 field and atomic operators read  $a_k(t)=\exp(\ri H_0t/\hbar)a_k\exp(-\ri
H_0t/\hbar)=a_k\exp(-\ri\omega_k t)$,
$\sigma_+^{\mathrm{ex}}(t)=\sigma_+^{\mathrm{ex}}\exp(\ri\Delta t)$, and
$\sigma_+^{\mathrm{in}}(t)=\sigma_+^{\mathrm{in}}\exp(\ri\omega_0 t)$. We
thus arrive at the final form of
the Hamiltonian
\begin{eqnarray}
H_{AF}(t)&=&\hbar\sum_kg_k\Big(\cos\kappa\left(\re^{\ri(\omega_0-\omega_k)t}a_k\sigma_+^{\mathrm{in}}+\re^{-\ri(\omega_0-\omega_k)t}a_k^\dagger\sigma_-^{\mathrm{in}}\right)
\nonumber\\
&&+\ri\sin\kappa
\big(\re^{\ri(\omega_0+\Delta-\omega_k)t}a_k\sigma_+^{\mathrm{ex}}\sigma_+^{\mathrm{in}}+\re^{\ri(\omega_0-\Delta-\omega_k)t}a_k\sigma_-^{\mathrm{ex}}\sigma_+^{\mathrm{in}}\nonumber\\
&&-\re^{-\ri(\omega_0+\Delta-\omega_k)t}a_k^\dagger\sigma_-^{\mathrm{ex}}\sigma_-^{\mathrm{in}}-\re^{-\ri(\omega_0-\Delta-\omega_k)t}a_k^\dagger\sigma_+^{\mathrm{ex}}\sigma_-^{\mathrm{in}} \big)\Big)\,.\label{Hfint}
\end{eqnarray}

\subsection{Internal dynamics --- spontaneous emission}
Let us first examine the process of spontaneous emission for the tunneling
atom. We write a general pure state of the entire system (atom+field) as
\begin{equation} \label{psi}
|\psi(t)\rangle=\sum_{\bn,\sigma=\pm,\mu=g,e}c_{\bn\sigma\mu}|\bn\sigma\mu\rangle\,,
\end{equation}
where
$|\bn\sigma\mu\rangle\equiv|\bn\rangle|\sigma\rangle_{\mathrm{ex}}|\mu\rangle_{\mathrm{in}}$,
$|\bn\rangle=\prod_k|n_k\rangle$ is a product state of all the field
modes, $n_k=0,1,2,\ldots$ denotes the occupation number of mode $k$,
and the sum over $\bn$ is over all sets $\{n_0,n_1,\ldots\}$,
$|\sigma\rangle_{\mathrm{ex}}$ denotes the atomic external state and
$|\mu\rangle_{\mathrm{in}}$ the atomic internal state . We start
with a general initial state without any photon, but with the atom
excited internally, and externally in an arbitrary pure state,
\begin{equation} \label{psi0}
|\psi(0)\rangle=c_{\bZero+e}(0)|\bZero +e\rangle+c_{\bZero-e}(0)|\bZero
 -e\rangle\,.
\end{equation}
Normalization imposes $|c_{\bZero+e}(t)|^2+|c_{\bZero-e}(t)|^2=1$. From the
Schr\"odinger equation in the interaction picture, $\ri \hbar
\frac{d}{dt}|\psi(t)\rangle=H_{AF}(t)|\psi(t)\rangle$, we obtain the
equations of motion for the relevant coefficients,
\begin{eqnarray}
\ri\dot{c}_{\bZero-e}&=&\sum_kg_k\,\re^{\ri(\omega_0-\omega_k)t}\left(\cos\kappa\,c_{1_k-g}+
\ri\sin\kappa\,\re^{-\ri\Delta t}c_{1_k+g}\right)\label{c1}\,,\\
\ri\dot{c}_{\bZero+e}&=&\sum_kg_k\,\re^{\ri(\omega_0-\omega_k)t}\left(\cos\kappa\,c_{1_k+g}+
\ri\sin\kappa\,\re^{\ri\Delta t}c_{1_k-g}\right)\label{c2}\,,\\
\ri\dot{c}_{1_k-g}&=&\sum_kg_k\,\re^{-\ri(\omega_0-\omega_k)t}\left(\cos\kappa\,c_{\bZero-e}-
\ri\sin\kappa\,\re^{-\ri\Delta t}c_{\bZero+e}\right)\label{c3}\,,\\
\ri\dot{c}_{1_k+g}&=&\sum_kg_k\,\re^{-\ri(\omega_0-\omega_k)t}\left(\cos\kappa\,c_{\bZero+e}-
\ri\sin\kappa\,\re^{\ri\Delta t}c_{\bZero-e}\right)\label{c4}\,,
\end{eqnarray} where the dot means derivative with respect to the
time $t$. We can formally integrate Eqs.~(\ref{c3},\ref{c4}) and
insert them into Eqs.~(\ref{c1},\ref{c2}). This leads to a closed
system of equations for the coefficients $c_{\bZero\pm e}$. In order
to avoid unnecessarily heavy notations, we focus momentarily on the
equation for $c_{\bZero - e}$, which can be compactly summarized as
\begin{eqnarray}
\dot{c}_{\bZero-e}&=&{\cal G}(\omega_0,c_{\bZero-e})+{\cal
  G}(\omega_0-\Delta,c_{\bZero-e})+{\cal G}_c(\omega_0,c_{\bZero-e})-{\cal
  G}_c(\omega_0-\Delta,c_{\bZero-e})\,,\label{dotc}\\
 {\cal
  G}(\omega_0,c_{\bZero-e})&\equiv&-\frac{1}{2}\sum_kg_k^2\int_0^tdt'\,\re^{\ri(\omega_0-\omega_k)(t-t')}c_{\bZero-e}(t')\,,\label{dc}\\
 {\cal
  G}_c(\omega_0,c_{\bZero-e})&\equiv&-\frac{1}{2}
  \sum_kg_k^2\cos
  2\kappa \int_0^tdt'\,\re^{\ri(\omega_0-\omega_k)(t-t')}c_{\bZero-e}(t')\,.
\end{eqnarray}
The equation for $c_{\bZero+e}$ can be found by substituting
$\Delta\to -\Delta$, and $c_{\bZero-e}\to c_{\bZero+e}$ in
Eq.~(\ref{dotc}). In principle, Eq.~(\ref{dotc}) contains two more
terms, one given by
\begin{equation} \label{}
\frac{\ri}{2}\sum_kg_k^2\sin 2\kappa
  \int_0^tdt'\,\re^{\ri(\omega_0-\omega_k)(t-t')}\re^{-\ri\Delta t'}c_{\bZero+e}(t')\,,
\end{equation}
and the other by an almost identical term with opposite sign and the
phase factor $\re^{-\ri\Delta t'}$ replaced by $\re^{-\ri\Delta
  t}$. However, we will find that the overwhelming part of the time
integrals comes from $t'\simeq t\pm b/c$, such that the two phases differ
only by $\Delta b/c$. This quantity represents a tunneling speed compared to
the speed of light, and has to be necessarily much smaller than
one, as otherwise the tunneling would have to be described in relativistic
terms. Indeed, even for very large tunneling splittings ($\sim$MHz) and well
separation ($\sim\mu$m, say), this ratio is of order $10^{-8}$ and thus
entirely 
negligible. Therefore, the two additional terms cancel to very good
approximation. \\

We replace the sum over $k$ by an integration in the limit of infinite
quantization volume $V$, use polar coordinates for $\bk$, and find
\begin{eqnarray}
{\cal G}_c(\omega_0,c_{\bZero-e})&=&{\cal
  G}_c^{(+)}(\omega_0,c_{\bZero-e})+{\cal
  G}_c^{(-)}(\omega_0,c_{\bZero-e}),\\
{\cal
  G}_c^{(\pm)}(\omega_0,c_{\bZero-e})&=&-\frac{\wp^2}{64\pi^2\epsilon_0\hbar
  c^3}\int_{-1}^1d\mu\,  h(\eta,\mu)\\
&&\times\int_0^t dt'\,\int_0^\infty d\omega\,\omega^3
  \re^{\ri(\pm\mu\frac{b}{c}-(t-t'))\omega}\re^{\ri
  \omega_0(t-t')}c_{\bZero-e}(t')\,,\\
h(\eta,\mu)&=&\left(\sin^2\eta+2\cos^2\eta+\mu^2(\sin^2\eta-2\cos^2\eta)\right)\,,
\end{eqnarray}
which is still exact, but makes Eq.~(\ref{dotc}) a complicated
differo-integral equation. In order to proceed, we resort to the
Wigner Weisskopf approximation \cite{Weisskopf30,Scully97}. This
amounts to realizing that the main contribution to the integral over
$\omega$ will arise from a narrow interval around $\omega=\omega_0$,
with a width of the order of the rate of spontaneous emission
$\Gamma$. The standard Wigner-Weisskopf vacuum spontaneous emission
rate reads
\begin{equation} \label{Gamma}
\Gamma(\omega_0)=\frac{\omega_0^3\wp^2}{3\pi\epsilon_0\hbar c^3}\,,
\end{equation}
which means that $\Gamma(\omega_0)/\omega_0=(4/3)\alpha (\omega_0
d/c)^2\ll 1$, where $\alpha\simeq 1/137$ is the fine-structure
constant, and $d=\wp/e_0$ the dipole length (dipole matrix element
divided by electron charge). The ratio $\omega_0 d/c$ must be much
smaller than one for the dipole approximation to hold. In our
problem, the rate of spontaneous emission will be hardly modified.
Thus, for all values of $\omega_0$, there is indeed a sharp peak of
width $\sim \Gamma$ in the integrand of the integral over $\omega$
(if one was to perform the integration over $t'$ first). The factor
$\omega^3$ varies only slowly on that scale, and can therefore be
pulled out of the integral. Moreover, the lower bound of the
$\omega$ integral can be extended to $-\infty$, such that the
$\omega$ integral leads to a Dirac-delta function $2\pi\delta(\pm\mu
b/c-(t-t'))$. The slight retardation $b/c$ corresponding to the time
of travel of a light signal between the two wells is important here
as it determines the $\mu$-interval that contributes, but can be
neglected in $c_{\bZero-e}$ after performing the integration over
$t'$, as $c_{\bZero-e}$ will evolve on a time scale $1/\Gamma\gg
b/c$. We thus arrive at
\begin{eqnarray}
{\cal
  G}_c(\omega_0,c_{\bZero-e})&=&-d(\eta,\beta)\frac{\Gamma(\omega_0)}{4}c_{\bZero-e}(t)\label{Gcfin}\\
d(\eta,\beta)&=&\frac{3}{4}\left((\sin^2\eta+2\cos^2\eta)\frac{\ri}{\beta}(1-\re^{\ri
  \beta})+
  (\sin^2\eta-2\cos^2\eta)\frac{-2\ri+\re^{\ri\beta}(2\ri+2\beta-\ri\beta^2)}{\beta^3}\right)
\nonumber\,,
\end{eqnarray}
with $\beta=\omega_0 b/c$.
The functional ${\cal G}(\omega_0,c_{\bZero-e})$ is obtained from
(\ref{Gcfin}) by taking the limit $\beta\to 0$ and thus gives
\begin{equation} \label{G}
{\cal
  G}(\omega_0,c_{\bZero-e})=-\frac{\Gamma(\omega_0)}{4}c_{\bZero-e}(t)\,,
\end{equation}
as $\lim_{\beta\to 0}d(\eta,\beta)=1$.
Note that the dependence on $\omega_0$ is both in $\Gamma(\omega_0)$ and in
 $\beta$.
Altogether we have
\begin{eqnarray}
\dot{c}_{\bZero\pm e}&=&-\frac{\Gamma_\pm}{2}c_{\bZero\pm e}\,,\\
\Gamma_\pm&=&\frac{1}{2}\Big(
\Gamma(\omega_0)+\Gamma(\omega_0\pm\Delta)
+\Gamma(\omega_0)d(\eta,\frac{\omega_0b}{c})-\Gamma(\omega_0\pm\Delta)d(\eta,\frac{(\omega_0\pm\Delta)b}{c})\Big)\,,
\label{dotc2}
\end{eqnarray}
with the obvious solution
\begin{equation} \label{c0pm}
c_{\bZero\pm e}(t)=c_{\bZero\pm e}(0)\re^{-\frac{\Gamma_\pm t}{2}}\,.
\end{equation}
It is convenient to express $\Gamma_\pm$ in terms of the standard
Wigner-Weisskopf rate $\Gamma(\omega_0)$ for a localized atom, and use the
dimensionless parameter $\delta=\Delta/\omega_0\ll 1$ introduced previously. We
then have
\begin{equation}
\frac{\Gamma_\pm}{\Gamma(\omega_0)}=\frac{1}{2}\Big(1+d(\eta,\beta)+(1\pm\delta)^3\big(1-d(\eta,\beta(1\pm\delta))\big)\Big) \,.\label{Gpm}
\end{equation}
For $\delta=0$ and $\beta\lesssim 3$ (i.e.~for finite $b$ a double--well
potential with infinite barrier), we are immediately led back to the
standard Wigner-Weisskopf rate for both initial states $|\bZero\pm
e\rangle$, $\Gamma_\pm=\Gamma(\omega_0)$. Thus, as long as there is
no tunneling, an arbitrary coherent superposition of the atom in the
right and in the left well does not change the spontaneous emission
at all. Similarly we get back $\Gamma_\pm=\Gamma(\omega_0)$ for
$\beta=0$.  For $\beta\gtrsim 1$, $d(\eta,\beta)$ vanishes as
$1/\beta$, and we have approximately
$\Gamma_\pm/\Gamma(\omega_0)=\frac{1}{2}\left(1+(1\pm\delta)^3\right)$.
This rate has the simple interpretation of arising from two
independent decay channels, from $|\pm e\rangle$ to $|\pm g\rangle$
or $|\mp g\rangle$. Transitions which do not change the external
state see the same level spacing $\omega_0$ as an atom without
tunneling degree of freedom, whereas a flip of the external state
changes the level spacing by $\pm\Delta$. Both transitions come with
the corresponding Wigner-Weisskopf rate, adjusted for the correct
overall level spacing, and the total rate is the average rate from
the two decay channels.

Contrary to the Wigner-Weisskopf case, the rates $\Gamma_\pm$ are in
general complex. For the decay of the probabilities $|c_{\bZero\pm
  e}|^2$, only the real part of
$d(\eta,\beta)$,
\begin{eqnarray}
\Re\, d(\eta,\beta)&=&\frac{3}{2}\Big(\left(
\sin^2\eta+2\cos^2\eta\right)\frac{\sin\beta}{\beta}\nonumber\\
&&+\left(\sin^2\eta-2\cos^2\eta\right)\frac{2\beta\cos\beta+(\beta^2-2)\sin\beta}{\beta^3}
\Big)\label{dr}
\end{eqnarray}
is relevant. This function equals one for $\beta=0$, depends only
slightly on $\eta$, and decays with some slight oscillations as a
function of $\beta$. However, for $\beta\gtrsim 1$, $\delta\ll 1$
(exponentially small overlap of wave functions), and for $\delta\sim
1$, $\beta\ll 1$ (small $\omega_0$), i.e.~in the regime attainable within
the two-level 
approximation made in this paper, we have that $\Gamma_\pm$ deviates
from $\Gamma(\omega_0)$ only very slightly and we will set
$\Gamma_\pm=\Gamma(\omega_0)\equiv\Gamma$ in the rest of the paper.

\subsection{External dynamics --- decoherence of the atomic tunneling}
The quantum mechanical expectation value for the average position $\langle
z(t)\rangle$ and thus the dynamics of the
external degree of freedom follow from
\begin{equation} \label{zt}
\langle z(t)\rangle=\frac{b}{2}\left(\rho_{+-}^{\mathrm{ex}}(t)\re^{-\ri \Delta
  t}+c.c.\right)\,,
\end{equation}
where the matrix element $\rho_{+-}^{\mathrm{ex}}(t)$ in the interaction
picture is
obtained from tracing out the internal and field degrees of freedom,
\begin{eqnarray}
\rho_{+-}^{\mathrm{ex}}(t)&=&c_{\bZero+e}(t)c_{\bZero-e}^*(t)+K(t)\label{rhopm}\\
K(t)&\equiv&\sum_k c_{1_k-g}^*(t)c_{1_k+g}(t)\,.
\end{eqnarray}
The first term in Eq.~(\ref{rhopm}) can be obtained immediately from
 Eq.~(\ref{c0pm}). In order to get $K(t)$,
we insert the solutions (\ref{c0pm}) for $c_{\bZero\pm e}(t)$ into
(\ref{c3},\ref{c4}). Direct integration of the latter equations gives
\begin{eqnarray}
c_{1_k\pm g}(t)&=&g_k\Big( c_{\bZero\pm e}(0)\cos\kappa
\frac{1-\re^{(-\ri(\omega_0-\omega_k)-\Gamma/2)t}}{(\omega_k-\omega_0)+\ri\Gamma/2}\nonumber\\
&&-i\, c_{\bZero\mp
e}(0)\sin\kappa\,\frac{1-\re^{(-\ri(\omega_0\mp\Delta-\omega_k)-\Gamma/2)t}}{(\omega_k-\omega_0\pm\Delta)+i\Gamma/2}\Big)\,.\label{c1k}
\end{eqnarray}
where we have set $\Gamma_\pm=\Gamma$ in accordance with the
previous section. It is possible to calculate the tunneling motion
including terms of order $\delta$, but the calculations are tedious
and the additional
 information gained compared to order zero in $\delta$ not
 illuminating. We therefore present here a simplified calculation which
 leads to a result valid up to corrections ${\cal O}(\delta)$.
The shifts $\pm\Delta$ in
 the denominator and exponent in Eq.~(\ref{c1k}) have to be kept. Otherwise,
 there will be obviously no tunneling at all. Formally, the need to keep
 $\Delta$ in the denominator and exponent of (\ref{c1k}) arises from the
 fact that there $\Delta$ has to
 be compared not to $\omega_0$ but to $\omega_0-\omega_k$, which can vanish,
 as $\omega_k$ varies from $0$ to $\infty$.

We restrict ourselves to real initial amplitudes $c_{\bZero\pm
e}(0)$ and proceed in a similar fashion as for the spontaneous
emission to evaluate the sum over all modes $k$ in
Eq.~(\ref{rhopm}), after inserting (\ref{c1k}). This leads to
\begin{eqnarray}
K(t)&=&\frac{\wp^2\,c_{\bZero-e}(0)c_{\bZero+e}(0)}{16\pi^2c^3\epsilon_0\hbar}\int_0^\infty
\,d\omega\,
\omega^3\int_{-1}^1d\mu\, h(\eta,\mu)\nonumber\\
&&\Bigg(\cos^2(\frac{\mu\tilde{\beta}}{2})\frac{\re^{-\Gamma
    t}-2\cos((\omega-\omega_0)t)\re^{-\Gamma
      t/2}+1}{(\omega-\omega_0)^2+\Gamma^2/4} \nonumber\\
&&+\sin^2(\frac{\mu\tilde{\beta}}{2})\frac{\re^{2\ri\Delta t}\re^{-\Gamma
    t}-2\re^{\ri\Delta t}\cos((\omega-\omega_0)t)\re^{-\Gamma
      t/2}+1}{(\omega-\omega_0+\Delta+\ri\frac{\Gamma}{2})(\omega-\omega_0-\Delta-\ri\frac{\Gamma}{2})}\Bigg)\nonumber\\
&\simeq&\frac{\wp^2\omega_0^3\,c_{\bZero-e}(0)c_{\bZero+e}(0)}{16\pi^2c^3\epsilon_0\hbar}\int_{-\infty}^\infty
\,d\omega\Bigg(a(\eta,\tilde{\beta})\frac{\re^{-\Gamma
    t}-2\cos((\omega-\omega_0)t)\re^{-\Gamma
      t/2}+1}{(\omega-\omega_0)^2+\Gamma^2/4} \nonumber\\
&&+\big(\frac{8}{3}-a(\eta,\tilde{\beta})\big)\frac{\re^{2\ri\Delta
t}\re^{-\Gamma
    t}-2\re^{\ri\Delta t}\cos((\omega-\omega_0)t)\re^{-\Gamma
      t/2}+1}{(\omega-\omega_0+\Delta+\ri\frac{\Gamma}{2})(\omega-\omega_0-\Delta-\ri\frac{\Gamma}{2})}\Bigg)\,,
\end{eqnarray}
with $\tilde{\beta}=\omega b/c$ and
\begin{equation} \label{a}
a(\eta,\beta)=(\sin^2\eta+2\cos^2\eta)\left(1+\frac{\sin\beta}{\beta}\right)+(\sin^2\eta-2\cos^2\eta)\left(\frac{1}{3}+\frac{2\beta\cos\beta+(\beta^2-2)\sin\beta}{\beta^3}\right)\,.
\end{equation}
We have made the same approximations as for the calculation of the rates of
spontaneous emission, i.e.~pulled out a factor $\omega_0^3$ from the
integral over $\omega$, and extended the lower bound of the integral to
$-\infty$. In principle the $\omega$--integral is UV divergent and would need
a cut--off (see \cite{Berman05} for a discussion of experimentally relevant
cut-offs). However, in order to conserve probability, the same
approximations as for the rates of spontaneous emission need to be made
here. Also, while there are two resonances now, $\omega_0\pm\Delta$, they
differ only at order $\delta$, so that at lowest order in $\delta$ it is
indeed enough to pull out $\omega_0^3$ from the integral. The remaining $\omega$--integral is performed by contour
integration, and we find the final result
\begin{eqnarray}
\langle
z(t)\rangle&=&b\,c_{\bZero-e}(0)c_{\bZero+e}(0)\Bigg[\re^{-\Gamma
  t}\cos(\Delta t)
+\frac{3}{8}\Big\{a(\eta,\beta)(1-\re^{-\Gamma t})\cos(\Delta t)\nonumber\\
&&+\left(\frac{8}{3}-a(\eta,\beta)\right)\frac{\gamma/2}{1+\gamma^2/4}\left(
(1+\re^{-\Gamma t})\sin(\Delta t)+\frac{\gamma}{2}(1-\re^{-\Gamma
t})\cos(\Delta t) \right)
\Big\}\Bigg]\nonumber\\
&&\times\left(1+{\cal O}(\delta)\right)\,,\label{ztfin}
\end{eqnarray}
where we have introduced $\gamma\equiv \Gamma/\Delta$.

Let us consider a few special cases of this general result. First of
all, Eq.~(\ref{ztfin}) shows that for $c_{\bZero-e}(0)=0$ or
$c_{\bZero+e}(0)=0$, $\langle z(t)\rangle=0$ for all $t$. This
corresponds to putting the atom externally into one of the two
energy eigenstates $|\pm\rangle$ of the uncoupled system, which are
symmetric with respect to $z=0$.  The two decay channels do not
introduce a position bias either, and therefore the atom stays on
average always at $z=0$. Tunneling with full amplitude needs an
initial preparation in the right or left well,
i.e.~$c_{\bZero-e}(0)c_{\bZero+e}(0)=\pm 1/2$, and we will therefore
assume from now on $c_{\bZero-e}(0)c_{\bZero+e}(0)=1/2$.

The limit $\beta\to 0$ leads with
$\lim_{\beta\to 0} a(\eta,\beta)=8/3$ immediately  to
$\langle z(t)\rangle=\frac{b}{2}\cos(\Delta t)$,
i.e.~undisturbed tunneling motion, as if the atom had no internal structure
at all. The physical reason for this is of course that the emitted photon
has in this case a wavelength much larger than the distance between the two
wells such
that it does not carry any information about the position of the
atom, and therefore no decoherence of the tunneling motion arises. \\

The case of $\beta\gtrsim 1$ is more subtle. At $\beta\simeq 3$ we
have $a(\eta,\beta)\simeq 4/3$ and hence
\begin{equation} \label{ztbetalarge}
\langle
z(t)\rangle\simeq\frac{b}{2}\Bigg[\left(\frac{1}{2}(1+\re^{-\Gamma
t})+\frac{\gamma^2/8(1-\re^{-\Gamma t})}{1+\gamma^2/4}
\right)\cos(\Delta t)+\frac{\gamma/4}{1+\gamma^2/4}(1+\re^{-\Gamma
  t})\sin(\Delta t)
\Bigg]\,,
\end{equation}
which for $t\gg \Gamma^{-1}$, i.e.~when a photon has certainly been emitted,
settles down to
\begin{eqnarray}
\langle
z(t)\rangle&\simeq&\frac{b}{2}\Bigg[\left(\frac{1}{2}+\frac{\gamma^2/8}{1+\gamma^2/4}
\right)\cos(\Delta t)+\frac{\gamma/4}{1+\gamma^2/4}\sin(\Delta t)\Bigg]\label{ztibl}\\
&\simeq&\frac{b}{2}A\cos(\Delta t+\varphi)\,,
\end{eqnarray}
with
\begin{equation} \label{A}
A=\sqrt{\frac{\gamma^2+1}{\gamma^2+4}}\,.
\end{equation}
As a consequence, after a period of initial damping, tunneling with
a finite amplitude $Ab/2$, which is in general reduced compared to
the full possible value $b/2$, and phase shift $\varphi$ persists.

 This is very much in contrast to standard decoherence scenarios of
 a particle
 tunneling through a potential barrier \cite{Ankerhold95}, where
 the continued
coupling to a heat-bath normally destroys all coherence (even though
exceptions are possible in other contexts, in particular for heat-baths with
 small cut-off
frequency, which can lead to incomplete decoherence as well
\cite{Braun02}). Here, the decoherence is switched off once the
photon is emitted, as the center--of--mass coordinate of the atom
does not couple directly to the electromagnetic modes. Furthermore,
the time at which the photon is emitted, plays a crucial role. If we
take $\gamma=\Gamma/\Delta\to\infty$ in Eq.~(\ref{A}), we find
$A=1$, i.e.~in spite of strong dissipation
 and short wavelength of the photon, there is no decoherence of the
 tunneling motion at all. The reason lies in the fact that the photon is
 emitted immediately after preparation of the atom in the right well,
 i.e.~at a time, when it is not in a coherent superposition of eigenstates
 of its
 center--of--mass position. Thus, no coherence can get destroyed, and since
 after the
 emission of the photon decoherence is switched off, tunneling proceeds in
 the ground state with full amplitude. One may also see the emission of the
 photon as a measurement process which should, for $\lambda\lesssim b$, project the atom either into the
 right or left well. But since the atom was prepared in the right well just
 before, one projects the state back into the right well, therefore keeping
 the coherence of the initial external state.

On the other hand, if we take $\gamma\to 0$ in Eq.~(\ref{A}), we
find that the amplitude of the tunneling motion in the long time
limit reduces to $A=1/2$. In the many runs of the experiment
necessary to verify
 Eq.~(\ref{ztfin}), the time when the photon is emitted is averaged over
 many tunneling periods, such that roughly speaking
in half the runs the atom is in a coherent superposition of eigenstates of its
center--of--mass position $z$, half of the time it is in an eigenstate of
 $z$. Therefore, it
is natural that on the average tunneling with half the full
 amplitude persists for $t\gg \Gamma^{-1}$.

\begin{figure}
\begin{center}
\includegraphics[width=.45\linewidth]{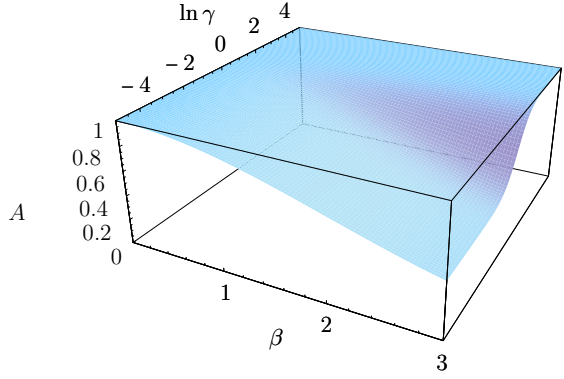}
\end{center}
\caption{(Color online) Amplitude of the tunneling motion for $t\gg \Gamma^{-1}$ for
  $\eta=0$ as function of $\beta$ and $\ln\gamma$.} \label{fig.A}
\end{figure}

\begin{figure}
\begin{center}
\includegraphics[width=.45\linewidth]{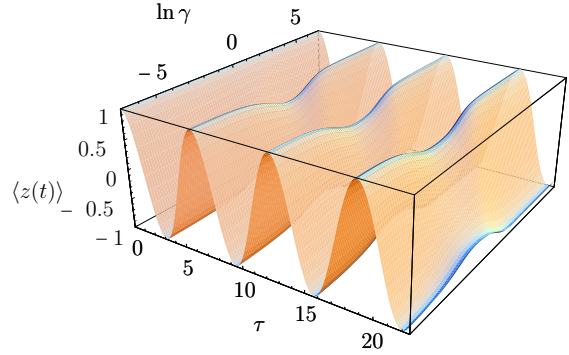}
\end{center}
\caption{(Color online) Tunneling motion $\langle z(t)\rangle$ in units of $b/2$ as
function of $\tau\equiv
  \Delta t$ and
  $\ln\gamma$ for $\beta=3$, $\eta=0$ with initial value $\langle
  z(0)\rangle=b/2$. } \label{fig.zt}
\end{figure}

The limits $\gamma\to 0$ and $t\to\infty$ do not commute, which is a
consequence of the factors $\exp(-\Gamma t)$ in $\langle
z(t)\rangle$. If we take the limit $\gamma\to 0$ already in
Eq.~(\ref{ztfin}), i.e.~without considering $t\to\infty$ first, we
get $\langle z(t)\rangle=\frac{b}{2}\cos (\Delta t)$. The atom
tunnels with full amplitude, i.e.~$A=1$, and shows no decoherence,
as it should be, of course. Therefore, in a finite time interval
starting with the preparation of the initial state, decoherence is
most effective in an intermediate regime, $\Gamma\sim 2\Delta$, when
a photon is likely to be emitted at the time when tunneling has
established a coherent superposition of $|R\rangle$ and $|L\rangle$.
For general $\eta,\gamma$ and large times, $t\gg \Gamma^{-1}$,
i.e.~after the damping has settled down, $\langle z(t)\rangle/(b/2)$
oscillates with an amplitude
\begin{equation} \label{Af}
A=\frac{1}{4}\sqrt{\frac{9\,a(\eta,\beta)^2+16\gamma^2}{4+\gamma^2}}\,,
\end{equation}
a function which we show in Fig.~\ref{fig.A}. This makes clear that
also at very small (but finite $\gamma$), $A$ reduces to 1/2 for
sufficiently large $\beta$. When plotting $\langle z(t)\rangle$ in a
fixed time interval as in Fig.~\ref{fig.zt}, the reduction of $A$ is
not visible yet at small values of $\gamma$.

\subsection{The electromagnetic field}
We now calculate the electromagnetic field in the limit of large
times, $t\gg \Gamma_\pm^{-1}$, such that the atom has emitted a
photon with certainty. We will approximate again
$\Gamma_+\simeq\Gamma_-\simeq\Gamma(\omega_0)\equiv\Gamma$ (we are
restricted to the regime $\omega_0\gg \Delta, \Gamma$). The
wave function of the entire system has then the form
\begin{equation}
\ket{\psi(\infty)}=\sum_{k}\big(c_{1_{k}-g}(\infty)\ket{1_{k}-g}
+c_{1_{k}+g}(\infty)\ket{1_{k}+g}\big)
\end{equation}
with
\begin{equation}
    c_{1_{k}\pm g}(\infty)=g_k\left(\frac{c_{\bZero \pm e}(0)\cos\kappa}{\omega_k-\omega_0+i\Gamma/2}
    -i\:\frac{c_{\bZero \mp e}(0)\sin\kappa}{\omega_k-\omega_0\pm\Delta+i\Gamma/2}\right)\label{c1kpm}
\end{equation}
(see (\ref{c1k})).  The first order
correlation function of the electric field $G^{(1)}(\vec{r},\vec{r};t,t)$
\cite{Scully97} is given for large times
 by
\begin{eqnarray}
G^{(1)}(\vec{r},\vec{r};t,t) & =&
\bra{\psi(\infty)}\bE^{(-)}(\vec{r},t).\bE^{(+)}(\vec{r},t)\ket{\psi(\infty)}=|\bI^+|^2+|\bI^-|^2\,,\\
\bI^\pm&=&\sum_{k}c_{1_{k}\pm
g}\bra{0}\bE^{(+)}(\vec{r},t)\ket{1_{k}}\label{Ipm}\,,
\end{eqnarray}
with the positive frequency electric field operator
$\bE^{(+)}(\vec{r},t)=\sum_{k}
    {\cal E}_k\beps_k a_k\re^{\ri (\bk.
    \br-\omega_k t)}$. The difference between $\bI^+$ and $\bI^-$ rests upon
$c_{1_k\pm g}$. We focus for the moment on $\bI^+$.  The
corresponding results for $\bI^-$ are obtained in a completely
analogous fashion. In fact, according to Eqs.~(\ref{c1kpm},\ref{Ipm})
one only needs to exchange $c_{\bZero + e}\leftrightarrow c_{\bZero
- e}$ and replace $\Delta$ by $-\Delta$ in the final result to
obtain $\bI^-$ from $\bI^+$. We choose the vector $\vec{r}$ to lie
in the $x-z$ plane, convert the sum over $k$ into an integral as
before,  and are thus led to
\begin{eqnarray}
\bI^+& =& -\frac{1}{16\pi^3\epsilon_0c^3}\int_0^\pi \sin\theta
d\theta \int_0^{2\pi} d\phi\int_0^{+\infty} d\omega \omega^3
\left(\dm-\vec{k}\frac{\vec{k}.\dm}{k^2}\right)e^{i(\vec{k}.\vec{r}-\omega t)}\nonumber\\
&&  \;\;\;\;\;\times \left(\frac{c_{\bZero +
e}(0)\cos\kappa}{\omega-\omega_0+i\Gamma/2}
    -i\:\frac{c_{\bZero -
    e}(0)\sin\kappa}{\omega-\omega_0+\Delta+i\Gamma/2}\right)\\
    & =& \bI_c^++\bI_s^+\,,
\end{eqnarray}
where $\vec{k}.\vec{r}=(\omega/c)(z\cos\theta+x\sin\theta\cos\phi)$.
The integral splits into two parts $\bI_c^+$, $\bI_s^+$ with
denominators $\omega-\omega_0+i\Gamma/2$ and
$\omega-\omega_0+\Delta+i\Gamma/2$. As in the Wigner-Weisskopf
theory of spontaneous emission, we assume that $\omega^3$ varies
little around $\omega=\omega_0$ (respectively
$\omega=\omega_0-\Delta$) so that we can replace $\omega^3$ by
$\omega_0^3$ (respectively $(\omega_0-\Delta)^3$) and extend the
lower limit of integration  to $-\infty$. Only the $x$ and
$z$-components of $\bI_c^+$ and $\bI_s^+$, denoted as $I_{c,\xi}^+$
$I_{s,\xi}^+$, $\xi\in\{x,z\}$, give a contribution (the
$y$-component is identically zero),
\begin{eqnarray}
I_{c,\xi}^+
 &=& -\frac{\omega_0^3\wp\,c_{\bZero +
e}(0)}{16\pi^3c^3\epsilon_0}\int_0^\pi \sin\theta d\theta
\int_0^{2\pi} d\phi\;f_\xi(\eta,\theta,\phi)\nonumber\\
&&\times\int_{-\infty}^{+\infty}
d\omega\frac{e^{i\omega(\frac{z}{c}\cos\theta+\frac{x}{c}\sin\theta\cos\phi
-t)}\cos\left(\frac{\omega
b}{2c}\cos\theta\right)}{\omega-\omega_0+i\Gamma/2}\,,\\
f_x(\eta,\theta,\phi)&=&[\sin\eta-\sin\theta\cos\phi(\sin\theta\cos\phi\sin\eta+\cos\theta\cos\eta)]\,,\\
f_z(\eta,\theta,\phi)&=&[\cos\eta-\cos\theta(\sin\theta\cos\phi\sin\eta+\cos\theta\cos\eta)]\,.
\end{eqnarray}

Writing $\cos\left(\frac{\omega b}{2c}\cos\theta\right)$ as a sum of
exponentials, the $\omega$-integral can be easily evaluated using
the contour method and leads to a Heaviside $\Theta$-function
$\Theta\Big(t-((z\pm b/2)\cos\theta+x\sin\theta\cos\phi)/c\Big)$ for
the positive (negative) frequency part of the
$\cos\left(\frac{\omega b}{2c}\cos\theta\right)$ function, which
illustrates the fact that the electric field cannot spread faster
than the speed of light. In order to avoid the complications which
arise from the angle dependence of the $\Theta$-function, we will
restrict ourselves to times $t>\frac{1}{c}(r+b/2)$ with
$r=\sqrt{x^2+z^2}$. We find
\begin{eqnarray}
I_{c,\xi}^+
 &=&i\frac{\omega_0^3\wp\,c_{\bZero +e}(0)}{8\pi^2c^3\epsilon_0}\int_0^\pi \sin\theta d\theta
\int_0^{2\pi} d\phi\;f_\xi(\eta,\theta,\phi)\\
&& \;\;\;\;\;\times
e^{i(\omega_0-i\Gamma/2)(\frac{z}{c}\cos\theta+\frac{x}{c}\sin\theta\cos\phi
-t)}\cos\Big(\frac{b}{2c}(\omega_0-i\Gamma/2)\cos\theta\Big)\,.\nonumber
\end{eqnarray}
Integration over the angle $\phi$ yields for the $x$-component
\begin{eqnarray}
I_{c,x}^+
 & =& i\frac{\omega_0^3\wp\,c_{\bZero+
e}(0)}{4\pi c^3\epsilon_0}\int_0^\pi \sin\theta d\theta\;
e^{i(\omega_0-i\Gamma/2)(\frac{z}{c}\cos\theta -t)}
\cos\Big(\frac{b}{2c}(\omega_0-i\Gamma/2)\cos\theta\Big)\nonumber\\
&& \times\Big\{\sin\eta\;
J_0\left(\frac{x}{c}(\omega_0-i\Gamma/2)\sin\theta\right)-i\sin\theta\cos\theta\cos\eta\;J_1\left(\frac{x}{c}(\omega_0-i\Gamma/2)\sin\theta\right)\nonumber\\
&& -\sin^2\theta\sin\eta\left[
\frac{J_1\left(\frac{x}{c}(\omega_0-i\Gamma/2)\sin\theta\right)}{\frac{x}{c}(\omega_0-i\Gamma/2)\sin\theta}
-J_2\left(\frac{x}{c}(\omega_0-i\Gamma/2)\sin\theta\right)\right]
\Big\}\,,
\end{eqnarray}
where $J_0$, $J_1$, and $J_2$ are Bessel functions.
A closed form of the type of the remaining $\theta$--integral has been found
very recently
by Neves et al.~\cite{Neves06}. Using their formula, we can evaluate the
$\theta$-integration analytically  and
find in the far-field region
\begin{eqnarray}\label{eqsinx}
I_{c,x}^+
 & =& i\frac{\omega_0^3\wp\,c_{\bZero+
e}(0)}{4\pi c^3\epsilon_0}\; e^{-i(\omega_0-i\Gamma/2)t}\nonumber
\\
&& \times\Big\{\sin\eta\;
\left[\frac{\sin R_+}{R_+}+\frac{\sin
 R_-}{R_-}\right]-\sin\eta\left[\frac{\sin^2\alpha_+\sin
 R_+}{R_+}+\frac{\sin^2\alpha_-\sin R_-}{R_-}\right]\nonumber\\
&&  -\frac{1}{2}\cos\eta\;\left[\frac{\sin (2\alpha_+)\sin
R_+}{R_+}+\frac{\sin (2\alpha_-)\sin R_-}{R_-}\right]
\Big\}+\mathcal{O}(1/R_\pm^2)\,,
\end{eqnarray}
with
\begin{equation}
\begin{aligned}
& R_\pm =\frac{(\omega_0-i\Gamma/2)}{c}\;r_\pm\,,\\
& r_\pm =\sqrt{x^2+(z\pm b/2)^2}\,,\\
& \tan\alpha_\pm =\frac{x}{z\pm b/2}\,.
\end{aligned}
\end{equation}
The expression for $I_{c,z}^+$ is obtained from $I_{c,x}^+$ by a
rotation of the coordinate system ($\eta\to\eta+\pi/2$,
$\alpha_\pm\to\alpha_\pm+\pi/2$), which amounts in
Eq.~\eqref{eqsinx} to exchanging $\sin\eta\leftrightarrow\cos\eta$
and replacing $\sin^2\alpha_{\pm}\rightarrow\cos\alpha_{\pm}^2$. In
the far-field, $\sin R_\pm\simeq e^{iR_\pm}/2i$, and the
expressions can be further simplified,
\begin{eqnarray}
I_{c,x}^+
 & =& \frac{\omega_0^3\wp\,c_{\bZero+e}(0)}{8\pi c^3\epsilon_0}\;
e^{-i(\omega_0-i\Gamma/2)t}\nonumber
\\
&& \times
\Bigg\{\frac{e^{iR_+}}{R_+}\sin(\eta-\alpha_+)\cos\alpha_+
+\frac{e^{iR_-}}{R_-}\sin(\eta-\alpha_-)\cos\alpha_-
\Bigg\}+\mathcal{O}(1/R_\pm^2)\,,\label{Icx2}
\end{eqnarray}

Similarly, we get for the second part $I_s^+$ of the integral
\begin{eqnarray}
I_{s,x}^+
 & =& -\frac{(\omega_0-\Delta)^3\wp\,c_{\bZero -
e}(0)}{8\pi c^3\epsilon_0}\;
e^{-i(\omega_0-\Delta-i\Gamma/2)t}\nonumber\\
&& \times
\Bigg\{\frac{e^{i\tilde{R}_+}}{\tilde{R}_+}\sin(\eta-\alpha_+)\cos\alpha_+
-\frac{e^{i\tilde{R}_-}}{\tilde{R}_-}\sin(\eta-\alpha_-)\cos\alpha_-
\Bigg\}+\mathcal{O}(1/\tilde{R}_\pm^2)\label{Isx2}
\end{eqnarray}
with
\begin{equation}
\tilde{R}_\pm =\frac{(\omega_0-\Delta-i\Gamma/2)}{c}\;r_\pm\,.
\end{equation}
The formulas for $I_{c,z}^+$ and $I_{s,z}^+$ are obtained by
changing the global sign and replacing
$\cos\alpha_{\pm}\rightarrow\sin\alpha_{\pm}$ where it appears
explicitly in (\ref{Icx2}) and (\ref{Isx2}), respectively . We have
\begin{equation}
\begin{aligned}
    I_{c,x}^++I_{s,x}^+ & \simeq\frac{\omega_0^2\wp }{8\pi c^2\epsilon_0}\frac{e^{-\frac{\Gamma}{2}(t-r/c)}}{r}\; e^{-i\omega_0t}\\
    & \;\;\;\;\;\times \Bigg\{ c_{\bZero+e}(0) \bigg[
    e^{i\frac{\omega_0}{c}r_+}\sin(\eta-\alpha_+)\cos\alpha_+
    + e^{i\frac{\omega_0}{c}r_-}\sin(\eta-\alpha_-)\cos\alpha_-
     \bigg] \\
    & \;\;\;\;\; - c_{\bZero -e}(0)\:e^{i\Delta t}\bigg[
    e^{i\frac{(\omega_0-\Delta)}{c}r_+}\sin(\eta-\alpha_+)\cos\alpha_+
    - e^{i\frac{(\omega_0-\Delta)}{c}r_-}\sin(\eta-\alpha_-)\cos\alpha_-
     \bigg] \Bigg\}\\
     & \;\;\;\;\; + \mathcal{O}(\Delta/\omega_0,\Gamma/\omega_0,1/r^2)\label{Icsx}\,,
\end{aligned}
\end{equation}
and again the corresponding expression for $I_{c,z}^++I_{s,z}^+$ can
be found from Eq.~(\ref{Icsx}) by just changing the global sign and
the explicitly printed factors $\cos\alpha_\pm$ into
$\sin\alpha_\pm$. We can summarize the results for both $\bI^+$ and
$\bI^-$ as
\begin{eqnarray}
   |\bI^\pm|^2 &
=&|I_{c,x}^\pm+I_{s,x}^\pm|^2+|I_{c,z}^\pm+I_{s,z}^\pm|^2\\
& \simeq& \frac{\omega_0^4\wp^2 }{64\pi^2 c^4\epsilon_0^2}\;
\frac{e^{-\Gamma(t-r/c)}}{r^2}\nonumber\\
&&\bigg\{\Big| c_{\bZero \pm e}(0) \Big[
    e^{i\frac{\omega_0}{c}\delta r}\sin(\eta-\alpha_+)\cos\alpha_+
    + \sin(\eta-\alpha_-)\cos\alpha_-
     \Big] \nonumber\\
    && - c_{\bZero \mp e}(0)\:e^{\pm i\Delta (t-r_-/c)}\Big[
    e^{i\frac{(\omega_0\mp\Delta) \delta r}{c}}\sin(\eta-\alpha_+)\cos\alpha_+
    - \sin(\eta-\alpha_-)\cos\alpha_-
     \Big] \Big|^2\nonumber\\
&&+\Big| c_{\bZero \pm e}(0) \Big[
    e^{i\frac{\omega_0}{c}\delta r}\sin(\eta-\alpha_+)\sin\alpha_+ +\sin(\eta-\alpha_-)\sin\alpha_-
     \Big]\nonumber\\
&& - c_{\bZero \mp e}(0)\:e^{\pm i\Delta (t-r_-/c)}\Big[
    e^{i\frac{(\omega_0\mp\Delta) \delta r}{c}}\sin(\eta-\alpha_+)\sin\alpha_+
    - \sin(\eta-\alpha_-)\sin\alpha_-
     \Big] \Big|^2 \nonumber\\
     && + \mathcal{O}(\Delta/\omega_0,\Gamma/\omega_0,1/r^2)\bigg\}\nonumber
\end{eqnarray}
with $\delta r=r_+-r_-$. In the limit $b\to 0$, we have $\delta
r=0$, $\alpha_\pm=\alpha$ with $\tan\alpha =x/z$, and we recover the
well-known result \cite{Scully94} (valid for $t>r/c$)
\begin{equation}
\begin{aligned}
G^{(1)}(\vec{r},\vec{r};t,t) & = \frac{\omega_0^4\wp^2 }{16\pi^2
c^4\epsilon_0^2}\;
\frac{e^{-\Gamma(t-r/c)}}{r^2}\;\sin^2(\eta-\alpha)\left(1+\mathcal{O}(\Delta/\omega_0,\Gamma/\omega_0,1/r^2)\right)
\end{aligned}
\end{equation}
where $\eta-\alpha$ is the angle between the dipole moment and the
observer.

Using $\delta r\simeq b\cos\alpha$, $\cos\alpha_\pm\simeq
\cos\alpha-\delta\alpha_\pm\sin\alpha$ and $\sin\alpha_\pm\simeq
\sin\alpha+\delta\alpha_\pm\cos\alpha$ with
$\delta\alpha_\pm=\alpha_\pm-\alpha\simeq \mp
\frac{b}{2r}\sin\alpha$ in the far-field, we get for an initial
delocalized state, $c_{\bZero \pm e}(0)=1$ and $c_{\bZero \mp
e}(0)=0$,
\begin{equation}
\begin{aligned}
G^{(1)}(\vec{r},\vec{r};t,t) & \simeq \frac{\omega_0^4\wp^2
}{16\pi^2 c^4\epsilon_0^2}\; \frac{e^{-\Gamma(t-r/c)}}{r^2}\;
    \sin^2(\eta-\alpha)\\
    & \;\;\;\;\; \times \left[1\pm
    \sin\Big(\frac{\Delta b}{2 c}\cos\alpha\Big)
    \sin\Big(\Big[\beta\pm\frac{\Delta b}{2 c}\Big]\cos\alpha\Big)
    \right]\\
    & \;\;\;\;\; \times \left(1+\mathcal{O}(\Delta/\omega_0,\Gamma/\omega_0,1/r^2)\right)\,,
\end{aligned}
\end{equation}
where upper and lower sign now refer to the initial condition.

If the atom is initially located in the right well,
$c_{\bZero +e}(0)=c_{\bZero -e}(0)=1/\sqrt{2}$ and we find
\begin{equation}\label{inter}
\begin{aligned}
G^{(1)}(\vec{r},\vec{r};t,t) & \simeq \frac{\omega_0^4\wp^2
}{16\pi^2 c^4\epsilon_0^2}\; \frac{e^{-\Gamma(t-r/c)}}{r^2}\;
    \sin^2(\eta-\alpha)\\
    & \;\;\;\;\; \times \left[1+
    \sin\Big(\frac{\Delta b}{2 c}\cos\alpha\Big)\bigg\{
    \cos\big(\beta\cos\alpha\big)\sin\Big(\frac{\Delta b}{2
    c}\cos\alpha\Big)\right. \\
    & \left. \;\;\;\;\;\;\;\;\;\;\;
    -\sin\Big(\Delta\left[t-\frac{r}{c}\right]\Big)
    \bigg(1+\cos\big(\beta\cos\alpha\big)\bigg)\bigg\}
    \right]\\
    & \;\;\;\;\; \times \left(1+\mathcal{O}(\Delta/\omega_0,\Gamma/\omega_0,1/r^2)\right)\,.
\end{aligned}
\end{equation}
In both cases the interference term is proportional to
$\sin\Big(\frac{\Delta b}{2 c}\cos\alpha\Big)$ which scales in the
non relativistic limit as $\Delta b/c$ and is thus extremely small.

\begin{figure}
\begin{center}
\includegraphics[width=.45\linewidth]{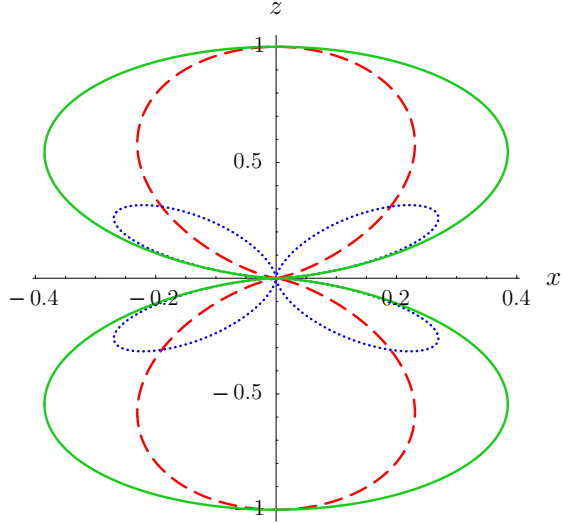}
\end{center}
\caption{(Color online) Polar plot of
$G^{(1)}_\pm(\vec{r},\vec{r};t,t)$ (red dashed and blue dotted
curves) and
$G^{(1)}(\vec{r},\vec{r};t,t)=G^{(1)}_+(\vec{r},\vec{r};t,t)+G^{(1)}_-(\vec{r},\vec{r};t,t)$
(green solid curve) for $\beta=3$, $\eta=0$, and an initially delocalized
state $c_{\bZero +e}(0)=1$.}
\label{postselect_b3eta0}
\end{figure}

The radiation from a classical oscillating dipole was examined
recently by Bolotovskii and Serov \cite{Bolotovskii07}. They predict
interference effects in the regime where $\beta\simeq 1$, but do not
provide any analytical results for the visibility of the
interference fringes. It appears, however, that a classically moving
dipole has to move a distance comparable to the wavelength during
the time $1/\omega_0$ if waves from different origins are to combine
in a remote location - otherwise the radiation pattern only follows
its source adiabatically. Indeed, it is easy to show that at order
zero in $\Delta b/c$ a classical oscillating dipole does not lead to
interference (where $\Delta$ means now the classical oscillation
frequency of the center--of--mass motion of the dipole). Therefore,
the absence of interference in Eq.~\eqref{inter} at lowest order in
$\Delta b/c$ agrees with the classical result. The initially
delocalized states $|\pm\rangle$ do not have a classical analogue.

However, it turns out that in the quantum case interference with
perfect visibility arises if the external state of the atom is
post-selected in the energy basis. For example, if only runs of the
experiments are taken into account where the atom is measured in
external state $|+\rangle$ before the photon is recorded, only
$|\bI^+|^2$ (but not $|\bI^-|^2$) contributes to
$G^{(1)}(\vec{r},\vec{r};t,t)$. We find in this case
$G^{(1)}_+(\vec{r},\vec{r};t,t)\propto
\cos^2(\frac{\beta}{2}\cos\alpha)\sin^2(\eta-\alpha)$ (subscript $+$
for post-selection in $|+\rangle$). The interference fringes for
post-selection in $|-\rangle$ are phase shifted, i.e.,
$G^{(1)}_-(\vec{r},\vec{r};t,t)\propto
\sin^2(\frac{\beta}{2}\cos\alpha)\sin^2(\eta-\alpha)$, such that in
the sum
$G^{(1)}(\vec{r},\vec{r};t,t)=G^{(1)}_+(\vec{r},\vec{r};t,t)+G^{(1)}_-(\vec{r},\vec{r};t,t)$
the interference terms compensate and only the dipole
characteristics $G^{(1)}(\vec{r},\vec{r};t,t)\propto
\sin^2(\eta-\alpha)$ is left (see Fig.~\ref{postselect_b3eta0} for
$\eta=0$). Note that for post-selection the interference pattern
becomes independent of $\Delta$ and should therefore be observable
even for small $\Delta$, as long as the two states can be reliably
distinguished.

\subsection{Experimental perspectives}\label{sec.exp}
Several requirements have to be met to observe the effects predicted
in this paper. First of all, one needs a double--well potential with
tunable well-to-well separation and barrier height in order to vary
$b$ and $\Delta$. This has been demonstrated with optical dipole
traps e.g.~in \cite{Shin04,Sebby06}, and on atom chips e.g.~in
\cite{Hinds01,Haensel01}. Secondly, we considered in our model the
same external potential for both internal states $|g\rangle$ and
$|e\rangle$, and these states should be coupled by a dipole
transition. It is by now well--known that this requirement can be
met for Cs, Yb, Sr, and possibly Mg and Ca atoms at certain
``magical wavelengths''  in optical traps
\cite{McKeever03,Katori03,Brusch06,Barber06}. Thirdly, we made the
two-level approximation for the external degree of freedom. It turns
out that in a typical double--well potential at the limit of the
Lamb-Dicke regime, $\beta\sim 3$ (i.e.\ $\lambda\sim 2b$) still
leads to a reasonable restriction of the dynamics to the two lowest
states. To show this, let us consider the usual quartic double--well
potential $V(x)=V_0(z^2-a^2)^2/a^4$ where $2a$ is the well-to-well
separation. Similar results are obtained for other shapes of the double--well
potential. The transition probability during emission of a photon
between energy eigenstates $|n\rangle$ and $|m\rangle$ of this
potential is given by $|\langle n| e^{ikz}|m
\rangle|^2$ \cite{CohenTannoudji01}. For Cs atoms and for the
parameters $V_0=0.23$ MHz and $a=\lambda/4$, where $\lambda=852.4$ nm  is 
the $6S_{1/2}\to 6P_{3/2}$ transition wavelength, this gives less than
$10\%$ transition 
probability out of the ground state doublet with a tunneling splitting
$\Delta\simeq 150$ Hz and 
$\beta\simeq 2.93$. For lighter atoms, such as Mg, tunnel splittings of the
order of kHz can be easily achieved for the same $\beta$. The theory thus
works well for $\beta\lesssim 3$ but not anymore for larger
values. Equation (\ref{Gamma}) implies that for a $\Delta$ in the kHz range,
$\Gamma\sim\Delta$ for a transition in the near-infrared ($\omega_0\sim
10^{13}$Hz), where a well-to-well separation of $\sim\mu$m leads to $\beta\sim
3$, which still keeps transition probabilities to higher vibrational states
at less than about 10$\%$.

 In any case, for optical traps the trap frequency and
thus the tunnel splitting are determined by the laser power and the
focusing (or the wavelength for optical lattices), and can therefore
be controlled independently of $\Gamma$ such that both regimes
$\Delta\gg \Gamma$ and $\Delta\ll \Gamma$ should be achievable. The
spontaneous emission rate $\Gamma$ can be varied over a large interval by
using a small static magnetic field to 
enable electrical dipole
transitions between hyperfine levels that would otherwise be
forbidden, due to admixture of small amplitudes of 
other hyperfine levels with allowed dipole transition
\cite{Taichenachev06,Mannervik96}. 

Cooling close to the ground state in a single well trap has  been
demonstrated and should work down to temperatures $k_B T<\hbar
\Delta$ if $\Delta$ is comparable to the single well vibrational
frequency \cite{McKeever03}. Finally, one needs to detect the
tunneling motion. That should be possible by optical imaging, i.e.
diffusion of laser light from another transition in the optical
regime with smaller wavelength than the well separation. Another
possibility might be using the atomic spin as a position meter
\cite{Haycock00}.

\section{Conclusions}
A two--level atom which can tunnel between the two wells of a
double--well potential allows for a host of interesting phenomena,
which we have studied systematically in this paper in the regime
$\delta=\Delta/\omega_0\ll 1$ (tunnel frequency much smaller than
the atomic transition frequency) and $\beta=\omega_0 b/c\lesssim 3$. Whereas
the spontaneous emission 
rate $\Gamma$ of the atom is only slightly modified by putting the
atom into a coherent symmetric superposition (ground state of the
external double--well potential) of the two states in the right and
left well, the tunneling of the atom and the properties of the
emitted light are altered more profoundly. The emission of a single
photon can cause decoherence of the tunneling motion, but only if
1.) the photon wavelength is not much larger than the distance
between the two potential wells, and 2.) spontaneous emission is not
too fast, i.e.~$\Gamma\lesssim \Delta$. For $\Gamma\gg \Delta$, the
photon is emitted even before the atom starts its tunneling motion,
i.e.~before a coherent superposition of eigenstates of the
center--of--mass coordinate of the atom is established. Hence,
despite strong coupling to the environment, the tunneling motion
does not suffer from decoherence at all in this regime. After the
emission of the photon no more decoherence takes place, and
tunneling will then continue with constant amplitude. For very slow
spontaneous emission, $\Gamma\ll \Delta$, the average amplitude of
the tunneling motion reduces by a factor 2. The electric field of
the emitted photon shows interference fringes, but their amplitude
is very small unless the external state of the atom is post-selected
in the energy basis, in which case interference fringes with perfect
visibility arise. This is very reminiscent of the results in
\cite{Schomerus02}, where it was predicted that interference fringes
from the scattering of a particle by a quantum scatterer disappear
in the limit where the kinetic energy $\epsilon\ll \Delta$, but can
be recovered by post-selecting the elastic scattering channel. The
effects predicted here should in principle be observable with modern
cold--atom technology.

{\em Acknowledgments:} We would like to thank CALMIP (Toulouse) for
the use of their computers. This work was supported by the Agence
National de la Recherche (ANR), project INFOSYSQQ, and the EC
IST-FET project EUROSQIP.

\bibliography{../../mybibs_bt}

\begin{thebibliography}{37}
\expandafter\ifx\csname natexlab\endcsname\relax\def\natexlab#1{#1}\fi
\expandafter\ifx\csname bibnamefont\endcsname\relax
  \def\bibnamefont#1{#1}\fi
\expandafter\ifx\csname bibfnamefont\endcsname\relax
  \def\bibfnamefont#1{#1}\fi
\expandafter\ifx\csname citenamefont\endcsname\relax
  \def\citenamefont#1{#1}\fi
\expandafter\ifx\csname url\endcsname\relax
  \def\url#1{\texttt{#1}}\fi
\expandafter\ifx\csname urlprefix\endcsname\relax\def\urlprefix{URL }\fi
\providecommand{\bibinfo}[2]{#2}
\providecommand{\eprint}[2][]{\url{#2}}

\bibitem[{\citenamefont{Lenz and Meystre}(1993)}]{Lenz93}
\bibinfo{author}{\bibfnamefont{G.}~\bibnamefont{Lenz}} \bibnamefont{and}
  \bibinfo{author}{\bibfnamefont{P.}~\bibnamefont{Meystre}},
  \bibinfo{journal}{Phys. Rev. A} \textbf{\bibinfo{volume}{48}},
  \bibinfo{pages}{3365} (\bibinfo{year}{1993}).

\bibitem[{\citenamefont{Eichmann et~al.}(1993)\citenamefont{Eichmann,
  Bergquist, Bollinger, Gilligan, Itano, Wineland, and Raizen}}]{Eichmann93}
\bibinfo{author}{\bibfnamefont{U.}~\bibnamefont{Eichmann}},
  \bibinfo{author}{\bibfnamefont{J.~C.} \bibnamefont{Bergquist}},
  \bibinfo{author}{\bibfnamefont{J.~J.} \bibnamefont{Bollinger}},
  \bibinfo{author}{\bibfnamefont{J.~M.} \bibnamefont{Gilligan}},
  \bibinfo{author}{\bibfnamefont{W.~M.} \bibnamefont{Itano}},
  \bibinfo{author}{\bibfnamefont{D.~J.} \bibnamefont{Wineland}},
  \bibnamefont{and} \bibinfo{author}{\bibfnamefont{M.~G.}
  \bibnamefont{Raizen}}, \bibinfo{journal}{Phys. Rev. Lett.}
  \textbf{\bibinfo{volume}{70}}, \bibinfo{pages}{2359} (\bibinfo{year}{1993}).

\bibitem[{\citenamefont{Itano et~al.}(1998)\citenamefont{Itano, Bergquist,
  Bollinger, Wineland, Eichmann, and Raizen}}]{Itano98}
\bibinfo{author}{\bibfnamefont{W.~M.} \bibnamefont{Itano}},
  \bibinfo{author}{\bibfnamefont{J.~C.} \bibnamefont{Bergquist}},
  \bibinfo{author}{\bibfnamefont{J.~J.} \bibnamefont{Bollinger}},
  \bibinfo{author}{\bibfnamefont{D.~J.} \bibnamefont{Wineland}},
  \bibinfo{author}{\bibfnamefont{U.}~\bibnamefont{Eichmann}}, \bibnamefont{and}
  \bibinfo{author}{\bibfnamefont{M.~G.} \bibnamefont{Raizen}},
  \bibinfo{journal}{Phys. Rev. A} \textbf{\bibinfo{volume}{57}},
  \bibinfo{pages}{4176} (\bibinfo{year}{1998}).

\bibitem[{\citenamefont{Agarwal et~al.}(2002)\citenamefont{Agarwal, von
  Zanthier, Skornia, and Walther}}]{Agarwal02}
\bibinfo{author}{\bibfnamefont{G.~S.} \bibnamefont{Agarwal}},
  \bibinfo{author}{\bibfnamefont{J.}~\bibnamefont{von Zanthier}},
  \bibinfo{author}{\bibfnamefont{C.}~\bibnamefont{Skornia}}, \bibnamefont{and}
  \bibinfo{author}{\bibfnamefont{H.}~\bibnamefont{Walther}},
  \bibinfo{journal}{Phys. Rev. A} \textbf{\bibinfo{volume}{65}},
  \bibinfo{pages}{053826} (\bibinfo{year}{2002}).

\bibitem[{\citenamefont{Feagin}(2006)}]{Feagin06}
\bibinfo{author}{\bibfnamefont{J.~M.} \bibnamefont{Feagin}},
  \bibinfo{journal}{Physical Review A (Atomic, Molecular, and Optical Physics)}
  \textbf{\bibinfo{volume}{73}}, \bibinfo{pages}{022108}
  (\bibinfo{year}{2006}).

\bibitem[{\citenamefont{Wickles and M\"uller}(2006)}]{Wickles06}
\bibinfo{author}{\bibfnamefont{C.}~\bibnamefont{Wickles}} \bibnamefont{and}
  \bibinfo{author}{\bibfnamefont{C.}~\bibnamefont{M\"uller}},
  \bibinfo{journal}{Europhys. Lett.} \textbf{\bibinfo{volume}{74}},
  \bibinfo{pages}{240} (\bibinfo{year}{2006}).

\bibitem[{\citenamefont{Sebby-Strabley
  et~al.}(2006)\citenamefont{Sebby-Strabley, Anderlini, Jessen, and
  Porto}}]{Sebby06}
\bibinfo{author}{\bibfnamefont{J.}~\bibnamefont{Sebby-Strabley}},
  \bibinfo{author}{\bibfnamefont{M.}~\bibnamefont{Anderlini}},
  \bibinfo{author}{\bibfnamefont{P.~S.} \bibnamefont{Jessen}},
  \bibnamefont{and} \bibinfo{author}{\bibfnamefont{J.~V.} \bibnamefont{Porto}},
  \bibinfo{journal}{Phys. Rev. A} \textbf{\bibinfo{volume}{73}},
  \bibinfo{pages}{033605} (\bibinfo{year}{2006}).

\bibitem[{\citenamefont{Shin et~al.}(2004)\citenamefont{Shin, Saba, Pasquini,
  Ketterle, Pritchard, and Leanhardt}}]{Shin04}
\bibinfo{author}{\bibfnamefont{Y.}~\bibnamefont{Shin}},
  \bibinfo{author}{\bibfnamefont{M.}~\bibnamefont{Saba}},
  \bibinfo{author}{\bibfnamefont{T.~A.} \bibnamefont{Pasquini}},
  \bibinfo{author}{\bibfnamefont{W.}~\bibnamefont{Ketterle}},
  \bibinfo{author}{\bibfnamefont{D.~E.} \bibnamefont{Pritchard}},
  \bibnamefont{and} \bibinfo{author}{\bibfnamefont{A.~E.}
  \bibnamefont{Leanhardt}}, \bibinfo{journal}{Phys. Rev. Lett.}
  \textbf{\bibinfo{volume}{92}}, \bibinfo{eid}{050405} (\bibinfo{year}{2004}).

\bibitem[{\citenamefont{H\"ansel et~al.}(2001)\citenamefont{H\"ansel, Reichel,
  Hommelhoff, and H\"ansch}}]{Haensel01}
\bibinfo{author}{\bibfnamefont{W.}~\bibnamefont{H\"ansel}},
  \bibinfo{author}{\bibfnamefont{J.}~\bibnamefont{Reichel}},
  \bibinfo{author}{\bibfnamefont{P.}~\bibnamefont{Hommelhoff}},
  \bibnamefont{and} \bibinfo{author}{\bibfnamefont{T.~W.}
  \bibnamefont{H\"ansch}}, \bibinfo{journal}{Phys. Rev. A}
  \textbf{\bibinfo{volume}{64}}, \bibinfo{pages}{063607}
  (\bibinfo{year}{2001}).

\bibitem[{\citenamefont{Treutlein et~al.}(2006)\citenamefont{Treutlein, Hansch,
  Reichel, Negretti, Cirone, and Calarco}}]{Treutlein06}
\bibinfo{author}{\bibfnamefont{P.}~\bibnamefont{Treutlein}},
  \bibinfo{author}{\bibfnamefont{T.~W.} \bibnamefont{Hansch}},
  \bibinfo{author}{\bibfnamefont{J.}~\bibnamefont{Reichel}},
  \bibinfo{author}{\bibfnamefont{A.}~\bibnamefont{Negretti}},
  \bibinfo{author}{\bibfnamefont{M.~A.} \bibnamefont{Cirone}},
  \bibnamefont{and} \bibinfo{author}{\bibfnamefont{T.}~\bibnamefont{Calarco}},
  \bibinfo{journal}{Physical Review A (Atomic, Molecular, and Optical Physics)}
  \textbf{\bibinfo{volume}{74}}, \bibinfo{pages}{022312}
  (\bibinfo{year}{2006}).

\bibitem[{\citenamefont{Carnal and Mlynek}(1991)}]{Carnal91}
\bibinfo{author}{\bibfnamefont{O.}~\bibnamefont{Carnal}} \bibnamefont{and}
  \bibinfo{author}{\bibfnamefont{J.}~\bibnamefont{Mlynek}},
  \bibinfo{journal}{Phys. Rev. Lett.} \textbf{\bibinfo{volume}{66}},
  \bibinfo{pages}{2689} (\bibinfo{year}{1991}).

\bibitem[{\citenamefont{Keith et~al.}(1991)\citenamefont{Keith, Ekstrom,
  Turchette, and Pritchard}}]{Keith91}
\bibinfo{author}{\bibfnamefont{D.~W.} \bibnamefont{Keith}},
  \bibinfo{author}{\bibfnamefont{C.~R.} \bibnamefont{Ekstrom}},
  \bibinfo{author}{\bibfnamefont{Q.~A.} \bibnamefont{Turchette}},
  \bibnamefont{and} \bibinfo{author}{\bibfnamefont{D.~E.}
  \bibnamefont{Pritchard}}, \bibinfo{journal}{Phys. Rev. Lett.}
  \textbf{\bibinfo{volume}{66}}, \bibinfo{pages}{2693} (\bibinfo{year}{1991}).

\bibitem[{\citenamefont{Hackerm\"uller
  et~al.}(2004)\citenamefont{Hackerm\"uller, Hornberger, Brezger, Zeilinger,
  and Arndt}}]{Hackermueller04}
\bibinfo{author}{\bibfnamefont{L.}~\bibnamefont{Hackerm\"uller}},
  \bibinfo{author}{\bibfnamefont{K.}~\bibnamefont{Hornberger}},
  \bibinfo{author}{\bibfnamefont{B.}~\bibnamefont{Brezger}},
  \bibinfo{author}{\bibfnamefont{A.}~\bibnamefont{Zeilinger}},
  \bibnamefont{and} \bibinfo{author}{\bibfnamefont{M.}~\bibnamefont{Arndt}},
  \bibinfo{journal}{Nature} \textbf{\bibinfo{volume}{427}},
  \bibinfo{pages}{711} (\bibinfo{year}{2004}).

\bibitem[{\citenamefont{Miffre et~al.}(2006)\citenamefont{Miffre, Jacquey,
  B{\"u}chner, Tr{\'e}nec, and Vigu{\'e}}}]{Miffre06}
\bibinfo{author}{\bibfnamefont{A.}~\bibnamefont{Miffre}},
  \bibinfo{author}{\bibfnamefont{M.}~\bibnamefont{Jacquey}},
  \bibinfo{author}{\bibfnamefont{M.}~\bibnamefont{B{\"u}chner}},
  \bibinfo{author}{\bibfnamefont{G.}~\bibnamefont{Tr{\'e}nec}},
  \bibnamefont{and}
  \bibinfo{author}{\bibfnamefont{J.}~\bibnamefont{Vigu{\'e}}},
  \bibinfo{journal}{Phys. Scr.} \textbf{\bibinfo{volume}{74}},
  \bibinfo{pages}{C15} (\bibinfo{year}{2006}).

\bibitem[{\citenamefont{Monroe et~al.}(1996)\citenamefont{Monroe, Meekho, King,
  and Wineland}}]{Monroe96}
\bibinfo{author}{\bibfnamefont{C.}~\bibnamefont{Monroe}},
  \bibinfo{author}{\bibfnamefont{D.~M.} \bibnamefont{Meekho}},
  \bibinfo{author}{\bibfnamefont{B.~E.} \bibnamefont{King}}, \bibnamefont{and}
  \bibinfo{author}{\bibfnamefont{D.~J.} \bibnamefont{Wineland}},
  \bibinfo{journal}{Science} \textbf{\bibinfo{volume}{272}},
  \bibinfo{pages}{1131} (\bibinfo{year}{1996}).

\bibitem[{\citenamefont{Cohen-Tannoudji
  et~al.}(1992)\citenamefont{Cohen-Tannoudji, Bardou, and Aspect}}]{Cohen92}
\bibinfo{author}{\bibfnamefont{C.}~\bibnamefont{Cohen-Tannoudji}},
  \bibinfo{author}{\bibfnamefont{F.}~\bibnamefont{Bardou}}, \bibnamefont{and}
  \bibinfo{author}{\bibfnamefont{A.}~\bibnamefont{Aspect}}
  (\bibinfo{publisher}{World Scientific}, \bibinfo{address}{Singapore},
  \bibinfo{year}{1992}), Laser Sepectroscopy X.

\bibitem[{\citenamefont{Rohrlich et~al.}(2006)\citenamefont{Rohrlich, Neiman,
  Japha, and Folman}}]{Rohrlich06}
\bibinfo{author}{\bibfnamefont{D.}~\bibnamefont{Rohrlich}},
  \bibinfo{author}{\bibfnamefont{Y.}~\bibnamefont{Neiman}},
  \bibinfo{author}{\bibfnamefont{Y.}~\bibnamefont{Japha}}, \bibnamefont{and}
  \bibinfo{author}{\bibfnamefont{R.}~\bibnamefont{Folman}},
  \bibinfo{journal}{Phys. Rev. Lett.} \textbf{\bibinfo{volume}{96}},
  \bibinfo{pages}{173601} (\bibinfo{year}{2006}).

\bibitem[{\citenamefont{Schomerus et~al.}(2002)\citenamefont{Schomerus, Noat,
  Dalibard, and Beenakker}}]{Schomerus02}
\bibinfo{author}{\bibfnamefont{H.}~\bibnamefont{Schomerus}},
  \bibinfo{author}{\bibfnamefont{Y.}~\bibnamefont{Noat}},
  \bibinfo{author}{\bibfnamefont{J.}~\bibnamefont{Dalibard}}, \bibnamefont{and}
  \bibinfo{author}{\bibfnamefont{C.~W.~J.} \bibnamefont{Beenakker}},
  \bibinfo{journal}{Europhys. Lett.} \textbf{\bibinfo{volume}{57}},
  \bibinfo{pages}{651} (\bibinfo{year}{2002}).

\bibitem[{\citenamefont{Japha and Kurizki}(1996)}]{Japha96}
\bibinfo{author}{\bibfnamefont{Y.}~\bibnamefont{Japha}} \bibnamefont{and}
  \bibinfo{author}{\bibfnamefont{G.}~\bibnamefont{Kurizki}},
  \bibinfo{journal}{Phys. Rev. Lett.} \textbf{\bibinfo{volume}{77}},
  \bibinfo{pages}{2909} (\bibinfo{year}{1996}).

\bibitem[{\citenamefont{Martin and Braun}()}]{Martin07}
\bibinfo{author}{\bibfnamefont{J.}~\bibnamefont{Martin}} \bibnamefont{and}
  \bibinfo{author}{\bibfnamefont{D.}~\bibnamefont{Braun}},
  \eprint{quant-ph/0704.0763}.

\bibitem[{\citenamefont{Weisskopf and Wigner}(1930)}]{Weisskopf30}
\bibinfo{author}{\bibfnamefont{V.}~\bibnamefont{Weisskopf}} \bibnamefont{and}
  \bibinfo{author}{\bibfnamefont{E.}~\bibnamefont{Wigner}},
  \bibinfo{journal}{Z. Phys.} \textbf{\bibinfo{volume}{63}},
  \bibinfo{pages}{54} (\bibinfo{year}{1930}).

\bibitem[{\citenamefont{Scully and Zubairy}(1997)}]{Scully97}
\bibinfo{author}{\bibfnamefont{M.}~\bibnamefont{Scully}} \bibnamefont{and}
  \bibinfo{author}{\bibfnamefont{M.}~\bibnamefont{Zubairy}},
  \emph{\bibinfo{title}{Quantum Optics}} (\bibinfo{publisher}{Cambridge
  University Press}, \bibinfo{address}{Cambridge, UK}, \bibinfo{year}{1997}).

\bibitem[{\citenamefont{Berman}(2005)}]{Berman05}
\bibinfo{author}{\bibfnamefont{P.~R.} \bibnamefont{Berman}},
  \bibinfo{journal}{Physical Review A (Atomic, Molecular, and Optical Physics)}
  \textbf{\bibinfo{volume}{72}}, \bibinfo{pages}{025804}
  (\bibinfo{year}{2005}).

\bibitem[{\citenamefont{Ankerhold et~al.}(1995)\citenamefont{Ankerhold,
  Grabert, and Ingold}}]{Ankerhold95}
\bibinfo{author}{\bibfnamefont{J.}~\bibnamefont{Ankerhold}},
  \bibinfo{author}{\bibfnamefont{H.}~\bibnamefont{Grabert}}, \bibnamefont{and}
  \bibinfo{author}{\bibfnamefont{G.-L.} \bibnamefont{Ingold}},
  \bibinfo{journal}{Phys. Rev. E} \textbf{\bibinfo{volume}{51}},
  \bibinfo{pages}{4267} (\bibinfo{year}{1995}).

\bibitem[{\citenamefont{Braun}(2002)}]{Braun02}
\bibinfo{author}{\bibfnamefont{D.}~\bibnamefont{Braun}},
  \bibinfo{journal}{Phys. Rev. Lett.} \textbf{\bibinfo{volume}{89}},
  \bibinfo{pages}{277901} (\bibinfo{year}{2002}).

\bibitem[{\citenamefont{Neves et~al.}(2006)\citenamefont{Neves, Padilha,
  Fontes, Rodriguez, Curz, Barbosa, and Cesar}}]{Neves06}
\bibinfo{author}{\bibfnamefont{A.~A.~R.} \bibnamefont{Neves}},
  \bibinfo{author}{\bibfnamefont{L.~A.} \bibnamefont{Padilha}},
  \bibinfo{author}{\bibfnamefont{A.}~\bibnamefont{Fontes}},
  \bibinfo{author}{\bibfnamefont{E.}~\bibnamefont{Rodriguez}},
  \bibinfo{author}{\bibfnamefont{C.~H.~B.} \bibnamefont{Curz}},
  \bibinfo{author}{\bibfnamefont{L.~C.} \bibnamefont{Barbosa}},
  \bibnamefont{and} \bibinfo{author}{\bibfnamefont{C.~L.} \bibnamefont{Cesar}},
  \bibinfo{journal}{J. Phys. A: Math. Gen.} \textbf{\bibinfo{volume}{39}},
  \bibinfo{pages}{L293} (\bibinfo{year}{2006}).

\bibitem[{\citenamefont{Scully et~al.}(1994)\citenamefont{Scully, Walther, and
  Schleich}}]{Scully94}
\bibinfo{author}{\bibfnamefont{M.~O.} \bibnamefont{Scully}},
  \bibinfo{author}{\bibfnamefont{H.}~\bibnamefont{Walther}}, \bibnamefont{and}
  \bibinfo{author}{\bibfnamefont{W.~P.} \bibnamefont{Schleich}},
  \bibinfo{journal}{Phys. Rev. A} \textbf{\bibinfo{volume}{49}},
  \bibinfo{pages}{1562} (\bibinfo{year}{1994}).

\bibitem[{\citenamefont{Bolotovskii and Serov}(2007)}]{Bolotovskii07}
\bibinfo{author}{\bibfnamefont{B.~M.} \bibnamefont{Bolotovskii}}
  \bibnamefont{and} \bibinfo{author}{\bibfnamefont{A.~V.} \bibnamefont{Serov}},
  \bibinfo{journal}{Bull. of the Lebedev Phys. Inst.}
  \textbf{\bibinfo{volume}{34}}, \bibinfo{pages}{73} (\bibinfo{year}{2007}).

\bibitem[{\citenamefont{Hinds et~al.}(2001)\citenamefont{Hinds, Vale, and
  Boshier}}]{Hinds01}
\bibinfo{author}{\bibfnamefont{E.~A.} \bibnamefont{Hinds}},
  \bibinfo{author}{\bibfnamefont{C.~J.} \bibnamefont{Vale}}, \bibnamefont{and}
  \bibinfo{author}{\bibfnamefont{M.~G.} \bibnamefont{Boshier}},
  \bibinfo{journal}{Phys. Rev. Lett.} \textbf{\bibinfo{volume}{86}},
  \bibinfo{pages}{1462} (\bibinfo{year}{2001}).

\bibitem[{\citenamefont{McKeever et~al.}(2003)\citenamefont{McKeever, Buck,
  Boozer, Kuzmich, N\"agerl, Stamper-Kurn, and Kimble}}]{McKeever03}
\bibinfo{author}{\bibfnamefont{J.}~\bibnamefont{McKeever}},
  \bibinfo{author}{\bibfnamefont{J.~R.} \bibnamefont{Buck}},
  \bibinfo{author}{\bibfnamefont{A.~D.} \bibnamefont{Boozer}},
  \bibinfo{author}{\bibfnamefont{A.}~\bibnamefont{Kuzmich}},
  \bibinfo{author}{\bibfnamefont{H.-C.} \bibnamefont{N\"agerl}},
  \bibinfo{author}{\bibfnamefont{D.~M.} \bibnamefont{Stamper-Kurn}},
  \bibnamefont{and} \bibinfo{author}{\bibfnamefont{H.~J.}
  \bibnamefont{Kimble}}, \bibinfo{journal}{Phys. Rev. Lett.}
  \textbf{\bibinfo{volume}{90}}, \bibinfo{pages}{133602}
  (\bibinfo{year}{2003}).

\bibitem[{\citenamefont{Katori et~al.}(2003)\citenamefont{Katori, Takamoto,
  Pal'chikov, and Ovsiannikov}}]{Katori03}
\bibinfo{author}{\bibfnamefont{H.}~\bibnamefont{Katori}},
  \bibinfo{author}{\bibfnamefont{M.}~\bibnamefont{Takamoto}},
  \bibinfo{author}{\bibfnamefont{V.~G.} \bibnamefont{Pal'chikov}},
  \bibnamefont{and} \bibinfo{author}{\bibfnamefont{V.~D.}
  \bibnamefont{Ovsiannikov}}, \bibinfo{journal}{Phys. Rev. Lett.}
  \textbf{\bibinfo{volume}{91}}, \bibinfo{pages}{173005}
  (\bibinfo{year}{2003}).

\bibitem[{\citenamefont{Brusch et~al.}(2006)\citenamefont{Brusch, Targat,
  Baillard, Fouche, and Lemonde}}]{Brusch06}
\bibinfo{author}{\bibfnamefont{A.}~\bibnamefont{Brusch}},
  \bibinfo{author}{\bibfnamefont{R.~L.} \bibnamefont{Targat}},
  \bibinfo{author}{\bibfnamefont{X.}~\bibnamefont{Baillard}},
  \bibinfo{author}{\bibfnamefont{M.}~\bibnamefont{Fouche}}, \bibnamefont{and}
  \bibinfo{author}{\bibfnamefont{P.}~\bibnamefont{Lemonde}},
  \bibinfo{journal}{Physical Review Letters} \textbf{\bibinfo{volume}{96}},
  \bibinfo{pages}{103003} (\bibinfo{year}{2006}).

\bibitem[{\citenamefont{Barber et~al.}(2006)\citenamefont{Barber, Hoyt, Oates,
  Hollberg, Taichenachev, and Yudin}}]{Barber06}
\bibinfo{author}{\bibfnamefont{Z.~W.} \bibnamefont{Barber}},
  \bibinfo{author}{\bibfnamefont{C.~W.} \bibnamefont{Hoyt}},
  \bibinfo{author}{\bibfnamefont{C.~W.} \bibnamefont{Oates}},
  \bibinfo{author}{\bibfnamefont{L.}~\bibnamefont{Hollberg}},
  \bibinfo{author}{\bibfnamefont{A.~V.} \bibnamefont{Taichenachev}},
  \bibnamefont{and} \bibinfo{author}{\bibfnamefont{V.~I.} \bibnamefont{Yudin}},
  \bibinfo{journal}{Phys. Rev. Lett.} \textbf{\bibinfo{volume}{96}},
  \bibinfo{pages}{083002} (\bibinfo{year}{2006}).

\bibitem[{\citenamefont{Cohen-Tannoudji
  et~al.}(2001)\citenamefont{Cohen-Tannoudji, Dupont-Roc, and
  Grynberg}}]{CohenTannoudji01}
\bibinfo{author}{\bibfnamefont{C.}~\bibnamefont{Cohen-Tannoudji}},
  \bibinfo{author}{\bibfnamefont{J.}~\bibnamefont{Dupont-Roc}},
  \bibnamefont{and} \bibinfo{author}{\bibfnamefont{G.}~\bibnamefont{Grynberg}},
  \emph{\bibinfo{title}{Processus d'interaction entre photons et atomes}}
  (\bibinfo{publisher}{EDP Sciences/CNRS \'Editions}, \bibinfo{address}{75005
  Paris, France}, \bibinfo{year}{2001}).

\bibitem[{\citenamefont{Taichenachev et~al.}(2006)\citenamefont{Taichenachev,
  Yudin, Oates, Hoyt, Barber, and Hollberg}}]{Taichenachev06}
\bibinfo{author}{\bibfnamefont{A.~V.} \bibnamefont{Taichenachev}},
  \bibinfo{author}{\bibfnamefont{V.~I.} \bibnamefont{Yudin}},
  \bibinfo{author}{\bibfnamefont{C.~W.} \bibnamefont{Oates}},
  \bibinfo{author}{\bibfnamefont{C.~W.} \bibnamefont{Hoyt}},
  \bibinfo{author}{\bibfnamefont{Z.~W.} \bibnamefont{Barber}},
  \bibnamefont{and} \bibinfo{author}{\bibfnamefont{L.}~\bibnamefont{Hollberg}},
  \bibinfo{journal}{Physical Review Letters} \textbf{\bibinfo{volume}{96}},
  \bibinfo{pages}{083001} (\bibinfo{year}{2006}).

\bibitem[{\citenamefont{Mannervik et~al.}(1996)\citenamefont{Mannervik,
  Brostr\"om, Lidberg, Norlin, and Royen}}]{Mannervik96}
\bibinfo{author}{\bibfnamefont{S.}~\bibnamefont{Mannervik}},
  \bibinfo{author}{\bibfnamefont{L.}~\bibnamefont{Brostr\"om}},
  \bibinfo{author}{\bibfnamefont{J.}~\bibnamefont{Lidberg}},
  \bibinfo{author}{\bibfnamefont{L.-O.} \bibnamefont{Norlin}},
  \bibnamefont{and} \bibinfo{author}{\bibfnamefont{P.}~\bibnamefont{Royen}},
  \bibinfo{journal}{Phys. Rev. Lett.} \textbf{\bibinfo{volume}{76}},
  \bibinfo{pages}{3675} (\bibinfo{year}{1996}).

\bibitem[{\citenamefont{Haycock et~al.}(2000)\citenamefont{Haycock, Alsing,
  Deutsch, Grondalski, and Jessen}}]{Haycock00}
\bibinfo{author}{\bibfnamefont{D.~L.} \bibnamefont{Haycock}},
  \bibinfo{author}{\bibfnamefont{P.~M.} \bibnamefont{Alsing}},
  \bibinfo{author}{\bibfnamefont{I.~H.} \bibnamefont{Deutsch}},
  \bibinfo{author}{\bibfnamefont{J.}~\bibnamefont{Grondalski}},
  \bibnamefont{and} \bibinfo{author}{\bibfnamefont{P.~S.}
  \bibnamefont{Jessen}}, \bibinfo{journal}{Phys. Rev. Lett.}
  \textbf{\bibinfo{volume}{85}}, \bibinfo{pages}{3365} (\bibinfo{year}{2000}).

\end{thebibliography}

\end{document}